\documentclass[10pt,aps,twocolumn,amsmath,amssymb,eqsecnum,prb]{revtex4-1}
\usepackage[colorlinks=false]{hyperref}
\newcommand{\cQ}{{\cal Q}}
\begin{document}

\title{Topological central charge from Berry curvature: gravitational anomalies in trial wavefunctions for
topological phases}
\author{Barry Bradlyn}
\email{barry.bradlyn@yale.edu}
\affiliation{Department of Physics, Yale University, P.O. Box 208120, New Haven, Connecticut 06520-8120,
USA}
\author{N. Read}
\date{April 2, 2015}
\affiliation{Department of Physics, Yale University, P.O. Box 208120, New Haven, Connecticut 06520-8120,
USA}

\begin{abstract}
We show that the topological central charge of a topological phase can be
directly accessed from the ground-state wavefunctions for a system on a surface as a Berry curvature
produced by adiabatic variation of the metric on the surface, at least up to addition of another
topological invariant that arises in some cases. For trial wavefunctions that are given by conformal
blocks (chiral correlation functions) in a conformal field theory (CFT), we carry out this calculation
analytically, using the hypothesis of generalized screening. The topological central charge is found to
be that of the underlying CFT used in the construction, as expected. The calculation makes use of the
gravitational anomaly in the chiral CFT. It is also shown that the Hall conductivity can be obtained
in an analogous way from the U($1$) gauge anomaly.
\end{abstract}

\maketitle

\section{Introduction}\label{intro}
It has long been appreciated that there exists a deep connection between topologically-protected
non-dissipative zero-frequency transport coefficients and Berry curvature. The archetypal example of
this is the quantized Hall conductivity in the integer and fractional quantum Hall effects, which can
be expressed as a Berry curvature via the Kubo formula\cite{TKNN,Niu1985}. Additionally, the Hall
viscosity---an analogous non-dissipative contribution to the viscosity tensor of a fluid---can be
expressed as a Berry curvature associated with adiabatic changes of the aspect ratio or metric tensor of
a system on a torus \cite{Avron1995,Levay1995,Read2009}, and is related\cite{Read2009,Read2011} to the
so-called shift in the number of flux in the ground state on a sphere\cite{Wen1992}. Moreover, these
properties are related to Chern-Simons terms in the effective (induced) action of the system (the first
Wen-Zee term \cite{Wen1992} in the case of Hall viscosity). The thermal Hall conductivity is related to
the topological central charge of the edge theory\cite{Read2000}, that is the difference of
the central charges of right and left moving modes on the edge; the topological central charge is
a topological property that takes the same value throughout a topological phase of matter. It has long
been hoped that the thermal Hall conductivity can be connected with the gravitational Chern-Simons term
\begin{equation}
S_{\rm GCS}\propto\int d^3x\, \widehat{\epsilon}^{\mu\nu\lambda}\left(\Gamma^\rho_{\mu\sigma}
\partial_{\nu}\Gamma^{\sigma}_{\lambda\rho}+\frac{2}{3}\Gamma^{\rho}_{\mu\sigma}
\Gamma^{\sigma}_{\nu\theta}\Gamma^{\theta}_{\lambda\rho}\right),
\end{equation}
which is expected to appear in the effective action with the topological
central charge as its coefficient\cite{Read2000}. Recently, however, it has been shown that the thermal
Hall conductivity comes entirely from the edge\cite{Stone2012,Bradlyn2014}; accessing the topological
central charge from the bulk requires more than just the Kubo formula for thermal conductivity.

In addition to Ref.\ \onlinecite{Bradlyn2014}, several other recent papers have discussed the form of
the effective action for fractional quantum Hall
states\cite{Abanov2014-0,Abanov2014,Gromov2015,Can2014-0,Can2014,Can2014-1,Cho2014,Gromov2014}, and
some of these address the various roles played by the central charge in particular. In contrast to
these works, in this paper we show how to obtain the topological central charge as a Berry curvature
associated with position-dependent changes in the spatial metric in the bulk of the system. This gives
us a method to compute the coefficient of the gravitational Chern-Simons term directly from the
ground-state wavefunction for a topological phase; in principle, this method can be applied as a
numerical diagnostic tool. (In some cases, the central charge appears in combination with other
topological invariants, not in isolation, as we discuss later.) To illustrate and validate the approach,
we apply it in detail to trial wavefunctions that can be viewed as conformal blocks\cite{Moore1991},
in cases in which they represent a topological phase. Using the gravitational anomaly in the chiral
conformal field theory (CFT), we calculate the Berry curvature. Thus we find that (as expected) the
topological central charge of these
states is equal to the central charge of the CFT used to construct the ground-state wavefunction in the
bulk. We note that this approach yields the topological central charge as a real number, and not only
up to addition of a multiple of $8$ or $24$, as can be obtained from the modular $S$-matrix (see e.g.\ 
Ref.\ \onlinecite{Zhang2012}) or from momentum polarization\cite{Qi2013}, respectively.

First, in Section~\ref{sec:actionberry} we review the connection between Berry phases and the low-energy
effective action when all dynamical degrees of freedom have been integrated out (we call this the induced
action; it describes the response to classical background fields). We pay particular attention to the
induced action relevant for describing fractional quantum Hall states at low energies. Next, in
Section~\ref{sec:prelim}, we introduce some tools to describe the geometry of space in the presence of
perturbations of the metric that we have to consider. In Section~\ref{sec:holfact}, we review the
construction of trial wavefunctions for quantum Hall states from conformal blocks\cite{Moore1991}, and
extend the construction to include perturbations of both the background vector potential and the metric
of space. The role of neutralizing background charges in the CFT, and their relation to the external
magnetic field, is brought to the forefront. Also in that Section, we introduce the notion of
a holomorphic factorization anomaly. Roughly speaking, this is the failure of a CFT correlator to
factorize into holomorphic and antiholomorphic functions of some background fields; both
a U($1$) gauge field and a metric perturbation cause such anomalies. Using these tools, in
Section~\ref{sec:berry} we calculate the Berry curvature that is found when the metric of space is
changed adiabatically; we show that in this way the topological central charge $c$ of the topological
phase of matter can be obtained from Berry curvature and is related to the two-dimensional gravitational
anomaly. In parallel with this, we also rederive the Hall conductivity and Hall viscosity from the same
point of view, thus connecting the Hall conductivity with the U($1$) gauge anomaly.
In Sec.\ \ref{sec:disc} we make some comments on the form of the results available from the
method, on some deficiencies of the Wen-Zee results\cite{Wen1992}, and on multicomponent states, 
before concluding.

\section{Berry curvature from the low-energy effective theory}\label{sec:actionberry}

To begin, we first review the connection between Berry phases and low-energy effective actions.
While fairly straightforward, this point of view has not received much attention in the recent literature.
Consider a system with action $S[\phi,\cQ]$, where $\phi$ stands for all the quantum fields describing
the microscopic or internal degrees of freedom, and we suppose that the system is coupled to some
external (classical) fields $\cQ$ that we can control. The propagator $U[\phi_i,\phi_f,\cQ]$ between
an initial state $\left|\phi_i\right>$  at time $t_i$ and a final state $\left|\phi_f\right>$ at time $t_f$
can be written as the path integral
\begin{equation}
U[\phi_i,\phi_f,\cQ]=\int\mathcal{D}\phi\, e^{iS[\phi,\cQ]}, \label{propeq}
\end{equation}
in which $\phi$ has initial values $\phi_i$, and final value $\phi_f$.
Let us assume that the action $S$ supports a topological (i.e.\ gapped) phase as a ground state, and that
the gap does not close as $\cQ$ is varied. (Here we will ignore edges, for example by putting our system on
a closed surface.) In this case, we can take $\left|\phi_i\right>$ to be the ground state, and let $\cQ$
vary adiabatically in time. In particular, let us consider a variation of $\cQ$ around some {\em closed}
path in parameter space, i.e.\ \begin{equation}
\cQ(t_i)=\cQ(t_f).
\end{equation}
Then, by the adiabatic theorem, we will find that
\begin{equation}
U[\phi_i,\phi_i,\cQ]=e^{i\Omega},
\end{equation}
where $\Omega$ is the sum of the Berry phase $\gamma_B$ and the dynamical phase
\begin{equation}
\gamma_{D}=\int_{t_i}^{t_f}E(t)dt,\label{gammaD}
\end{equation}
where $E(t)$ is the ground state energy at time $t$.
Using the definition
\begin{equation}
iS_{\rm eff}[\cQ]=\ln\oint\mathcal{D}\phi e^{iS[\phi,\cQ]}
\end{equation}
of the effective or induced action, we see immediately that
\begin{equation}
\Omega=S_{\rm eff}[\cQ], \label{actionphaserel}
\end{equation}
that is, the change of phase of the ground state wavefunction accompanying an adiabatic change of
parameters around a closed path is given by the effective action evaluated on the path. (Here, for
simplicity, we assumed that there is a non-degenerate ground state; in the more general case in which
the ground state is degenerate, the Berry phase factor is replaced by a unitary matrix. However, in the
calculations considered in this paper, such a matrix reduces to a phase factor times the identity matrix.)

There are two sorts of terms that may appear in the effective action for a gapped phase\cite{Bradlyn2014}.
The first are locally-invariant terms, which can be written as integrals of local expressions that are
scalars under all local symmetries of the microscopic theory. Under adiabatic changes of the external
fields, the time integral of these terms depend on the rate at which the path in parameter space is
traversed. Hence these terms contribute to the dynamical phase $\gamma_D$ (in fact, these locally-invariant
terms are exactly the contributions to the ground state energy density of the system, and hence take the
form of Eq.\ (\ref{gammaD}) explicitly\cite{Bradlyn2014}).

The low energy action may also contain Chern-Simons terms. These terms are integrals
of local expressions that, rather than being invariant, change by total derivatives under the action of
local symmetries. These terms have two very important properties. First, as long as there is a gap in the
energy spectrum above the ground state everywhere in spacetime, their coefficients cannot be made
position dependent (doing so would destroy the symmetry of the induced action). Consequently, these
coefficients are topological properties, which within a topological phase do not change under changes 
of microscopic parameters. Second, and more directly
relevant for us, when it is non-zero, the \emph{value} of a Chern-Simons action as the external fields
are varied along a closed path does not depend on the rate at which the fields are varied, because
the integrand contains a part first-order in time derivatives. Thus, they contribute directly to the
\emph{Berry} phase $\gamma_B$.

Using Eq.\ (\ref{actionphaserel}), we can extract the Berry \emph{curvature} associated with a given
variation of external fields once we know the induced action, by a simple application of Stokes's
theorem to a small loop in $\cQ$ space. In this work, we will be interested in applying this idea to
the low-energy effective action for quantum Hall states. For the precise formulation, we use the
non-relativistic set-up of Ref.\ \onlinecite{Bradlyn2014} (for general differential geometry, see e.g.\
Ref.\ \onlinecite{Carroll2004}). Briefly, this involves a choice of a frame of
$d+1$ vectors with components $e^\mu_\alpha$ varying differentiably in spacetime; $\mu$ is an ambient
spacetime index, $\mu=0$, \ldots, $d$, and $\alpha$ is an internal index (for the members of the frame)
with the same range. The dual set of one-forms $e^\alpha_\mu$ obeys $e^\alpha_\mu
e^\mu_\beta=\delta^\alpha_\beta$ and $e^\alpha_\mu e^\nu_\alpha=\delta^\nu_\mu$; either the vectors or
the one-forms are referred to as the vielbeins. There is a Christoffel connection $\Gamma$ in spacetime
which is used to form covariant derivatives,
and we impose rotation invariance by letting it hold locally in the internal space indices $\alpha=a=1$,
\ldots, $d$; this involves the introduction of a spin connection
$\omega_{\mu\hphantom{\alpha}\beta}^{\hphantom{\mu}\alpha}$ which is zero for $\alpha$ or $\beta=0$.
We impose covariant constancy of the vielbein,
\begin{equation}
\nabla_\mu e_\nu^\alpha\equiv\partial_\mu
e_\nu^\alpha+\omega_{\mu\hphantom{\alpha}\beta}^{\hphantom{\mu}\alpha}e_\nu^\beta-
\Gamma_{\hphantom{\lambda}\mu\nu}^{\lambda}e_\lambda^\alpha=0.
\end{equation}
This allows us to express the Christoffel symbols as
\begin{equation}
\Gamma_{\hphantom{\lambda}\mu\nu}^{\lambda}=e^\lambda_\alpha\partial_\mu
e_\nu^\alpha+\omega_{\mu\hphantom{a}b}^{\hphantom{\mu}a}e^\lambda_ae^b_\nu.
\label{Christomegrel}
\end{equation}
Note that we do not assume the Christoffel symbols are symmetric on their lower indices, which means
the spacetime geometry could possess torsion.
For space dimension $d=2$ of interest here, the spin connection reduces to an internal scalar
\begin{equation}
\omega_\mu=\frac{1}{2}\epsilon_a^{\hphantom{a}b}\omega_{\mu\hphantom{a}b}^{\hphantom{\mu}a},
\end{equation}
as in the approach of Ref.\ \onlinecite{Wen1992}. Finally, in addition to the spacetime geometry, we have
a background electromagnetic field with potential [U($1$) connection] $A_\mu$, corresponding to the fact
that particle number is conserved. Except when otherwise noted, this is the only conservation law that 
we will assume in our system, apart from those for energy, momentum, and orbital angular momentum.

In terms of these external background fields, we can write down the most general induced action at
low order in derivatives. The Chern-Simons terms of interest
are\cite{Wen1992,Read2000,Abanov2014-0,Abanov2014,Bradlyn2014,Can2014,Can2014-1} (we drop the 
locally-invariant terms),
\begin{align}
S_{\rm eff}=&\frac{\nu}{4\pi}\int d^3x\,\widehat{\epsilon}^{\mu\nu\lambda}\left( A_\mu\partial_\nu
A_\lambda+2\bar{s}\omega_\mu\partial_\nu A_\lambda+
\overline{s^2}\omega_\mu\partial_\nu\omega_\lambda\right)\nonumber \\
&+\frac{c}{96\pi}\int d^3x\, \widehat{\epsilon}^{\mu\nu\lambda}\left(\Gamma^\rho_{\mu\sigma}
\partial_{\nu}\Gamma^{\sigma}_{\lambda\rho}+\frac{2}{3}\Gamma^{\rho}_{\mu\sigma}
\Gamma^{\sigma}_{\nu\theta}\Gamma^{\theta}_{\lambda\rho}\right),\label{Seff}
\end{align}
where $A$ is the electromagnetic potential or connection, $\omega$ is a non-relativistic $2+1$ dimensional
spin connection, and $\Gamma$ is the Christoffel connection. Here $\nu$ is the filling factor (no
confusion with the indices $\nu$ should occur), $\bar{s}$ is the mean orbital spin per particle,
$\overline{s^2}$ is the mean-squared orbital spin per particle, and $c$ is the topological central charge
of the edge theory. The first term here is the U($1$) Chern-Simons term. The next two terms are,
respectively, the first and second Wen-Zee terms. The last is the gravitational Chern-Simons term; the
fact that the coefficient is the topological central charge, which was defined in the previous section as
a property of the edge, will be discussed further in Section \ref{sec:disc}. Note that
$\widehat{\epsilon}^{\mu\nu\lambda}$ is the totally-antisymmetric epsilon {\em symbol}, not tensor,
defined by $\widehat{\epsilon}^{012}=1$.

In view of the relation (\ref{Christomegrel}), the gravitational Chern-Simons term,
\begin{equation}
S_\mathrm{GCS}=\frac{c}{96\pi}\int d^3x\, \widehat{\epsilon}^{\mu\nu\lambda}\left(\Gamma^\rho_{\mu\sigma}
\partial_{\nu}\Gamma^{\sigma}_{\lambda\rho}+\frac{2}{3}\Gamma^{\rho}_{\mu\sigma}
\Gamma^{\sigma}_{\nu\theta}\Gamma^{\theta}_{\lambda\rho}\right), \label{gcs}
\end{equation}
resembles the second Wen-Zee term. Indeed, substituting that relation into this expression, we find
after some algebra (essentially as in Refs.\ \onlinecite{Bardeen1984,AG-Pietra-Moore})
\begin{align}
S_{\mathrm{GCS}}=&-\frac{c}{48\pi}\int
d^3x\,\widehat{\epsilon}^{\mu\nu\lambda}\omega_\mu\partial_\nu\omega_\lambda\nonumber \\
&+\frac{c}{288\pi}\int d^3x\, \widehat{\epsilon}^{\mu\nu\lambda}e^\sigma_\beta(\partial_\mu e_\sigma^\alpha)
(\partial_\nu e_\rho^\beta)(\partial_\lambda e_\alpha^\rho).\label{gcsexp}
\end{align}
Finally, using
\begin{equation}
\partial_\nu\left(e_\mu^\alpha e^\mu_\beta\right)=0,
\end{equation}
the last term in Eq.\ (\ref{gcsexp}) can be expressed as
\begin{align}
&\widehat{\epsilon}^{\mu\nu\lambda}e^\sigma_\beta(\partial_\mu e_\sigma^\alpha)(\partial_\nu
e_\rho^\beta)(\partial_\lambda e_\alpha^\rho)\nonumber\\
&\qquad=\widehat{\epsilon}^{\mu\nu\lambda}\left(e^\alpha_\rho\partial_\mu
e^\rho_\beta\right)\left(e^\beta_\sigma\partial_\nu e^\sigma_\gamma\right)
\left(e^\gamma_\theta\partial_\lambda e^\theta_\alpha\right).
\end{align}
Written in this form, the second term in Eq.\ (\ref{gcsexp}) is a topological invariant (times the
topological central charge $c$), essentially the winding number of the
map defined by the vielbein from the spacetime manifold to the general linear group $GL(3,\mathbb{R})$,
and so does not vary under small variations of the vielbeins. (For manifolds
with boundary, there is a variation on the boundary).
We emphasize that while this result is equivalent to that presented in Ref.\ \onlinecite{Gromov2014},
we derived it here without any constraints on the torsion of the connection. Nonetheless, from this point
on we set the reduced torsion\cite{Bradlyn2014} to zero, as it plays no role.

Hence the gravitational Chern-Simons term can be re-written up to boundary contributions as
\begin{equation}
S_{\rm GCS}=-\frac{c}{48\pi}\int d^3x\,\widehat{\epsilon}^{\mu\nu\lambda}
\omega_\mu\partial_\nu\omega_\lambda,
\end{equation}
and thus, {\em as far as the bulk of the system is concerned}, it can be combined with the second Wen-Zee
term. The resulting term contains the coefficient
\begin{equation}
c_{\rm app}=c-12\nu\overline{s^2}
\end{equation}
which we term the ``apparent central charge.'' It has appeared before in special
cases\cite{Klevtsov2014, Can2014-0, Abanov2014,Can2014}; in particular for the $\nu=1/Q$ Laughlin state,
it gives $c_{\rm app}=1-3Q$ when the expected values $c=1$ and $\overline{s^2}=Q^2/4$ are used.

We see that in the induced action there are terms in which the explicit derivative is with respect to time,
and the indices on the connections $A_\mu$ or $\omega_\mu$ are spatial; these terms determine the Berry
phase for a loop and hence the Berry curvature for those types of variation of the external fields for
quantum Hall systems. Conversely, if we can compute from the ground-state wavefunction the Berry curvature
associated with some perturbations of the background fields, we can calculate the coefficients in Eq.\
(\ref{Seff}) directly; this will be carried out for some trial states in this paper. We note that the
Berry curvature can also be termed the ``symplectic form'' for the background fields in question, and
is similar to what occurs in relation to phase space in classical mechanics.

To illustrate this procedure, let us consider the U($1$) Chern-Simons action for $A_\mu$. We take
$A_0=0$ fixed for all time, while $A_1$ and $A_2$ evolve in time around some closed path in function
space. We can then write the Chern-Simons action as
\begin{align}
S_{CS}=\frac{\nu}{4\pi}\int d^2x \int dt&\left[A_2(\mathbf{x},t)\partial_0A_1(\mathbf{x},t)\right.
\nonumber \\
&\left.-A_1(\mathbf{x},t)\partial_0A_2(\mathbf{x},t)\right].
\end{align}
Now, examining the above, we note that $dt\partial_0A_\mu(\mathbf{x},t)$ is independent of the
parametrization $t$ of the path in function space. In fact, this combination is simply $
dA_\mu(\mathbf{x})$, the differential of $A_\mu(\mathbf{x})$ along the path at any $\bf x$, with $t$ used
as the path parameter. We may then identify the integral above with
\begin{equation}
S_{CS}=\int d^2x \int dA_\mu(\mathbf{x})\mathcal{A}^\mu(\mathbf{x}),
\end{equation}
where $\mu$ is summed from $1$ to $2$, and
\begin{align}
\mathcal{A}^1&=\frac{\nu}{4\pi}A_2, \nonumber \\
\mathcal{A}^2&=-\frac{\nu}{4\pi}A_1.
\end{align}
We can view $\mathcal{A}^\mu=\delta S_{CS}/\delta \partial_0 A_\mu$ as the functional derivative of
the action with respect to the time derivative $\partial_0 A_\mu$; it plays the role of a functional
Berry connection. By a further functional derivative, we can form the functional Berry curvature
\begin{align}
\mathcal{F}_{A_1,A_2}=\frac{\delta \mathcal{A}^2(\mathbf{x})}{\delta A_1(\mathbf{y})}
-\frac{\delta \mathcal{A}^1(\mathbf{x})}{\delta
A_2(\mathbf{y})}
\end{align}
($\mathcal{F}_{A_2,A_1}=-\mathcal{F}_{A_1,A_2}$), which here gives
\begin{equation}
\mathcal{F}_{A_1,A_2}=-\frac{\nu}{2\pi}\delta(\mathbf{x}-\mathbf{y}).\label{u1berr}
\end{equation}
Then applying Stokes's theorem in parameter space we obtain
\begin{equation}
S_{CS}=\int d^2x\int d^2y\int\int
dA_1(\mathbf{x})dA_2(\mathbf{y})\mathcal{F}_{A_1,A_2}(\mathbf{x},\mathbf{y}).
\end{equation}
(This involves a double integral over parameters, say $t$, $t'$, that cover the interior of the closed path
in function space.)

One can carry out a very similar calculation in the simpler case of $A_\mu$ constant in space on a $2$-torus
(equivalent to a twisted boundary condition), and the result is directly related to writing the Hall
conductivity as a Berry curvature or Chern number in the space of boundary conditions\cite{Niu1985}.
On the other hand, if in the same geometry we consider $A_\mu=\partial_\mu \Lambda$ ($\mu=1$, $2$) and
$A_0=0$ globally, with $\Lambda$ a globally defined function, so that the space components are those of
a pure gauge, the U($1$) Chern-Simons term for a closed loop in $\Lambda({\bf x})$-space vanishes. This
may raise some concerns as to whether we can in fact obtain the central charge as a Berry curvature in
this way from the gravitational Chern-Simons term using variations of the spatial metric. Nonetheless it
turns out that we can, as we will see.

\section{Differential-geometric preliminaries}\label{sec:prelim}

The results of the preceding section tell us how in principle to compute the coefficients of the
Chern-Simons terms in the effective action, including the topological central charge: we should find the
ground state in the presence of a perturbation of the background fields, and by varying that along a closed
path, the Berry curvature. We will carry out such a calculation for conformal-block wavefunctions, but
first we explain some further techniques for describing the background geometry, in particular the spatial
metric. These tools could be useful in calculations for general wavefunctions as well.

In the non-relativistic formalism of Ref.\ \onlinecite{Bradlyn2014}, the metric is
$g_{\mu\nu}=e_\mu^ae_\nu^b\eta_{ab}$,
where $\eta_{ab}=\delta_{ab}$ is the standard metric on the internal space indices only; thus $g_{\mu\nu}$
is degenerate. From this point on, we further assume that the timelike vielbein is trivial:
$e_\mu^0=\delta_\mu^0$, and similarly for $e^\mu_0$. With this convention we can view the indices $\mu$,
$\nu$, \ldots, as running over $1$, $2$ only; we adopt this convention when dealing with the
two-dimensional spatial geometry from here on. (In Ref.\ \onlinecite{Bradlyn2014}, $i$, $j$, \ldots,
were used, but in this paper we reserve these for particle labels.) Then the metric can be viewed
as $g_{\mu\nu}$, the spatial metric in each time slice.

We assume that the space at each time is an orientable surface without boundary; at some points it will be
convenient to assume the surface is a torus, but usually it can be any surface. We assume that we have
fixed a choice of an atlas of coordinate charts on the surface\cite{Carroll2004}, and use $x^1$, $x^2$
for the coordinates
in one chart (region). For the purposes of CFT, complex coordinates are useful, but the complex structure
involved is not unique, even given the fixed coordinate system. First, an {\em almost-complex structure} on
an orientable even-dimensional manifold is a (real) tensor field ${\cal J}^\mu_\nu$ satisfying ${\cal
J}^\mu_\lambda
{\cal J}^\lambda_\nu=-\delta^\mu_\nu$ (see Ref.\ \onlinecite{gsw}). In two dimensions, and if ${\cal J}$
varies sufficiently smoothly with position (which we will
tacitly assume to hold), it is always possible locally to find a system of {\em holomorphic} complex
coordinates $F$, $\bar{F}$ such that in $F$, $\bar{F}$ components ${\cal J}$ takes the form
\begin{align}
{\cal J}=\left(\begin{array}{cc}
i & 0  \\
0 & -i \end{array}\right).
\end{align}
$F$ is unique modulo conformal transformations that replace $F$ by a holomorphic function
of $F$. We can find an atlas of such local coordinate charts\cite{Carroll2004} related by local
conformal mappings in the
overlap regions. A {\em complex structure} is defined as the equivalence class of such atlases modulo
globally-defined conformal mappings; thus in two dimensions, any globally-defined ${\cal J}$ determines
a complex structure. On our standard coordinate chart, a particular almost-complex structure,
which we call the standard one, is determined by the standard complex coordinates $z=x^1+ix^2$,
$\bar{z}=x^1-ix^2$. In these coordinates we define complex components for vector fields and one-forms by
\begin{align}
B^z&=B^1+iB^2,\;B^{\bar{z}}=B^1-iB^2,\\
A_z&=\frac{1}{2}(A_1-iA_2),\; A_{\bar{z}}=\frac{1}{2}(A_1+iA_2);
\end{align}
for such components in the standard coordinates, we frequently omit the indices $z$, $\bar{z}$, and write 
for example $A=A_z$, $\bar{A}=A_{\bar{z}}$. Similarly, we write $\partial$ for 
$\partial_z=\frac{1}{2}(\partial_1-i\partial_2)$ and likewise for $\bar{\partial}$.

The general almost-complex structure ${\cal J}$ can alternatively be described using the standard one and a
position-dependent
{\em Beltrami differential} $\mu=\mu^z_{\bar{z}}$, which does not transform as a tensor (no confusion of
the Beltrami differential with a spacetime index should occur). Then a holomorphic
coordinate $F$ is defined by the property that its differential $dF$ is proportional to $dz+\mu d\bar{z}$
at each point, with $|\mu|<1$. This means that $F$ is a solution to the Beltrami equation (see e.g.\ Ref.\
\onlinecite{Bers1972})
\begin{equation}
\bar{\partial}F=\mu\partial F; \label{beltramieqn}
\end{equation}
this equation always has solutions for $F$. We note that
\begin{equation}
\frac{\partial}{\partial
\bar{F}}=\frac{1}{(1-|\mu|^2)\bar\partial \bar{F}}\left(\bar\partial-\mu\partial\right),
\end{equation}
such that $\partial F/\partial\bar{F}=0$, $\partial\bar{F}/\partial\bar{F}=1$; the first condition is
equivalent to the Beltrami equation.

If we change from our coordinates $z$, $\bar{z}$ to another set $\zeta$, $\bar{\zeta}$ (which are
functions of $z$, $\bar{z}$), but leave the almost-complex structure unchanged, then $dF$ is
unchanged, but $\mu$ is replaced by $\mu^\zeta_{\bar{\zeta}}$, with
\begin{align}
\mu^\zeta_{\bar{\zeta}}&=\frac{\frac{\partial z}{\partial\bar{\zeta}}+\mu\frac{\partial
\bar{z}}{\partial\bar{\zeta}}}{\frac{\partial z}{\partial{\zeta}}+\mu\frac{\partial
\bar{z}}{\partial{\zeta}}}\\
&=\frac{\mu\partial\zeta-\bar\partial\zeta}{\bar\partial\bar\zeta-\mu\partial\bar\zeta}.
\end{align}
This shows that $\mu$ does not transform as a tensor, and that a holomorphic coordinate system can be
viewed as a choice $\zeta=F$ such that $\mu^F_{\bar{F}}=0$. If we begin with the zero Beltrami, $\mu=0$,
and make an infinitesimal coordinate transformation $\zeta=z+f$, where $f$ is small, then in the $\zeta$
coordinates $\mu^\zeta_{\bar{\zeta}}=-\bar\partial f$ to first order.

Sometimes it is instead convenient to say that $dF$ is proportional to
$dx^1+\tau dx^2$ at each point, ${\rm Im}\,\tau>0$ (with $\tau$ depending on position). It is easy to
show that
\begin{equation}
\mu=\frac{i-\tau}{i+\tau}
\end{equation}
(this expression already appeared in Ref.\ \onlinecite{Read2011}).
As an example of distinct complex structures, consider the torus described in the standard coordinates as
the square region $0\leq x^1\leq L$, $0\leq x^2\leq L$ with periodic boundary conditions. If the
almost-complex structure is described in these coordinates by a position-independent $\tau$, and we define
$F=x^1+\tau x^2$, then in the
complex $F$ plane the region becomes a parallelogram, and the ratio of the two sides adjacent to the origin
(i.e.\ of the two distinct complex displacements along the sides in the complex $F$ plane) is $\tau$.
This is a standard way of describing the torus, used in the theory of elliptic functions (and in
Refs.\ \onlinecite{Read2009,Read2011}). Torii with different constant $\tau$ have inequivalent
complex structures (they cannot be related by global conformal maps), and for the torus (a surface of
genus one) the single complex parameter $\tau$ is the only parameter needed to determine the complex
structure\cite{Bers1972,gsw,Friedan1987}. For compact surfaces
of genus ${\cal G}>1$, complex structures are parametrized by $3{\cal G}-3$ complex
parameters (so-called moduli), while for the sphere (${\cal G}=0$), there is a unique complex structure.

Any spatial line element on the surface, $ds^2=g_{\mu\nu}dx^\mu dx^\nu$ in the standard coordinates,
can be written as\cite{Friedan1987}
\begin{equation}
ds^2=e^\Phi\left|dz+\mu d\bar{z}\right|^2\label{linelem}
\end{equation}
for some choice of $\mu$.
Here $e^{\Phi(z,\bar{z})}$ is the so-called conformal factor ($\Phi$ is real, and called the
Liouville field). Just as $\mu$ is not a tensor, $\Phi$ does not transform as a scalar: in coordinates
$\zeta(z,\bar{z})$, it becomes
\begin{eqnarray}
\Phi'&=&\Phi + \ln \left|\frac{\partial z}{\partial\zeta}+ \mu \frac{\partial
\bar{z}}{\partial\zeta}\right|^2\\
&=&\Phi-\ln\left|\partial\zeta+\mu_{\bar{\zeta}}^\zeta \partial\bar{\zeta}\right|^2.
\end{eqnarray}
The line element Eq.\ (\ref{linelem}) corresponds to a metric tensor $g$ which in complex coordinates
has components
\begin{align}
g_{zz}&= e^\Phi\bar{\mu}, \label{metric1} \\
g_{z\bar{z}}&=\frac{e^\Phi}{2}\left(1+\left|\mu\right|^2\right), \label{metric3}
\end{align}
and $g_{\bar{z}\bar{z}}=\overline{g_{zz}}$, $g_{\bar{z}z}=g_{z\bar{z}}$.
Let us notice that
\begin{equation}
\sqrt{\det\,g}=e^{\Phi}\left(1-|\mu|^2\right).\label{detgform}
\end{equation}
Later, in Sec.\ \ref{sec:berry}, it will be convenient to assume that $\det g=1$, but in general we will
not assume
this, and so $\mu$ and $\Phi$ are independent variables. We note that in holomorphic coordinates, the
line element becomes $ds^2=e^\Phi|dF|^2$, where $\Phi$ is not the same one as in the standard coordinates.
A nice example is the case of a sphere. In the quantum Hall literature, wavefunctions on the sphere are
commonly written in stereographic coordinates, which are holomorphic coordinates $F=z$ covering all of the
sphere except the south pole at $|z|\to\infty$. In those coordinates, $\mu=0$ and the rotation- [i.e.\
SO($3$)-] invariant metric is determined by
\begin{equation}
e^{\Phi(z,\bar{z})}=\frac{1}{[1+|z|^2/(4{\cal R}^2)]^2},
\end{equation}
where $\cal R$ is the radius of the sphere (see e.g.\ Ref.\ \onlinecite{Read2008}).

Returning to the general case, we can associate with the metric (\ref{metric1}), (\ref{metric3})
a ``canonical'' choice of vielbeins.
We can introduce internal complex components for internal vector and one-form fields as
\begin{equation}
\begin{array}{ll}
e^+=e^1+ie^2,&e^-=e^1-ie^2, \\
f_+=\frac{1}{2}(f_1-if_2),&f_-=\frac{1}{2}(f_1+if_2).
\end{array}
\end{equation}
With these conventions we can rewrite the line element $ds^2$
in terms of the vielbeins $e_\nu^\pm$, defined by
\begin{align}
e^+_\nu dx^\nu&=e^{\Phi/2}(dz+\mu d\bar{z}), \nonumber \\
e^-_\nu dx^\nu&=e^{\Phi/2}(d\bar{z}+\bar{\mu}dz),
\end{align}
as $ds^2= |e^+_\nu dx^\nu|^2$.
From the formalism of Ref.\ \onlinecite{Bradlyn2014}, the spin connection $\omega_\mu$ (see Sec.\
\ref{sec:actionberry}) in $z$, $\bar{z}$ space components is
\begin{align}
\omega_z&=\frac{i}{1-|\mu|^2}\left(\frac{1+|\mu|^2}{2}\partial
\Phi-\bar{\mu}\bar{\partial}\Phi+\bar{\mu}\partial\mu-\bar{\partial}\bar{\mu}\right),\nonumber\\
\omega_{\bar{z}}&=\overline{\omega_z}.\label{spinconndef}
\end{align}
It will also be useful later to know that
\begin{equation}
\omega_{\bar{z}}-\mu\omega_z=i\left[\partial
\mu-\frac{1}{2}\left(\bar\partial\Phi-\mu\partial\Phi\right)\right],
\label{spinconnF}
\end{equation}
and this is proportional to $\omega_{\bar{F}}$ at each point (as for $\partial/\partial\bar{F}$ above).
We should notice that if we construct the ``canonical'' combination $e^+_\nu dx^\nu$ in another coordinate
system, then it changes by a phase, the local rotation of the coordinates. This is unconventional, as a
combination $e^a_\nu dx^\nu$ should be invariant under coordinate transformations, but transform by
rotation on the internal space index $a$ under an internal rotation. The construction of the ``canonical''
vielbeins involves a choice of gauge for internal rotations that is related to the ambient space
coordinates, such that there now {\em is} a change of phase induced by a coordinate transformation when
this gauge choice is used. This induces an inhomogeneous term in the transformation of $\omega_\mu$ also.

A further geometric structure arises because we consider charged particles in a magnetic field; the field
is described by a vector potential $A_\mu$ on the surface (after choosing a gauge). A wavefunction for such
a particle (it is sufficient here to consider a single-particle wavefunction; the generalization to many
particles is performed in the usual way) transforms as a scalar under coordinate transformations, and by
a phase factor under a gauge transformation. We can define lowest Landau level (LLL) wavefunctions $\psi$
in our geometry analogously to their usual form, as functions that in terms of holomorphic coordinates $F$
are annihilated by the covariant derivative with respect to $\bar{F}$. This condition reduces to
\begin{equation}
(\bar{\partial}-\mu\partial - i\bar{A}+i\mu A)\psi=0.\label{vecholo}
\end{equation}
For $\mu=0$, this reduces further to the standard form often used in the quantum Hall effect
(we are assuming that the magnetic field strength $B({\bf x})=\partial_1A_2-\partial_2A_1$ is positive
everywhere). In the coordinate patch considered, the equation has solutions of the form
of a ``particular'' solution, a function of $F$ and $\bar{F}$ (or $z$ and $\bar{z}$), times
any function holomorphic in $F$. (These must be patched together using coordinate and gauge
transformations to obtain global solutions; on a compact surface, the space of solutions is finite
dimensional.) Usually the LLL functions are defined as the lowest-energy states
for the single-particle Hamiltonian $H_0=-g^{\mu\nu}(\nabla_\mu-iA_\mu)(\nabla_\nu-iA_\nu)/(2m_p)$,
where $m_p$ is the mass of the particle. With the non-trivial geometry, these states may not all be
degenerate, and may not be the same as we have defined. But by addition of a suitable potential term
to $H_0$ [namely, $-B({\bf x})/(2m_p\sqrt{\det g})$], the LLL
states defined here can be obtained as zero-energy eigenstates (when studying this,
it is convenient to use holomorphic coordinates). As we study
topological properties of a gapped phase of matter, and because if a topological phase occurs in the LLL
for such a Hamiltonian it can presumably be connected continuously with the same phase when it occurs for
the more conventional Hamiltonian, we are free to use such a definition and consider states in which
all particles are in the LLL as defined here.

\section{Conformal block wavefunctions and holomorphic factorization}\label{sec:holfact}

Our goal is to construct the ground states of a system in a topological phase on a surface with a
perturbation in the background metric, and also, both for the sake of completeness and for comparison,
with a perturbation in the background vector potential. We employ the conformal block construction of
Moore and Read to achieve this end. The details of the construction can be found in
Refs.~\onlinecite{Moore1991,Read2009}; we first summarize the main points of those works here and introduce
the related issues of holomorphic factorization and anomalies, which arise when we introduce perturbations
in the background fields. These subjects are explained in detail in the following subsections, first in
the case of the U($1$) gauge potential, then in the case of a metric perturbation.

\subsection{Basic construction and issues arising}

We begin with the case of a system forming a disk in the plane with the standard complex coordinates,
without perturbations of the background fields. We represent the (non-normalized) wavefunction
$\psi_e(z_1,z_2,....z_N)$ for a quantum Hall state at filling factor $\nu$
as a chiral correlator\cite{byb} (conformal block)
\begin{equation}
\lim_{\alpha\rightarrow 0}\left< \prod_{i=1}^N a(z_i)\prod_{j=1}^{N/\alpha}\mathcal{O}_{-\alpha
/\sqrt{\nu}}(w_j)\right>_0.
\end{equation}
Here the expectation $\langle\cdots\rangle_0$ is taken in the vacuum of the CFT without background field
perturbations, $i=1$ to $N$ labels the $N$ particles, and
$a(z)$ is the operator in the CFT that represents a physical particle (the electron or other underlying
particle). The CFT is the product of two other CFTs. The first is the theory of a scalar field $\varphi$
with action
\begin{equation}
S=\frac{1}{8\pi}\int d^2x \left(\nabla\varphi\right)^2.
\end{equation}
This theory has a U(1) current, which is $J=-i\partial \varphi(z)$ and a corresponding expression
for $\overline{J}$, and this current corresponds to the conserved ``charge'' or particle number, so this
CFT is called the charge sector.
The second theory is the ``statistics sector'', which can be used to obtain non-Abelian
quantum Hall states (for example, the Moore-Read state was obtained with a statistics sector given by
a free Majorana fermion theory\cite{Moore1991}). The particle operator is then
defined as the product of a chiral vertex operator
\begin{equation}
\mathcal{O}_{1/\sqrt{\nu}}(z)=e^{i\varphi(z)/\sqrt{\nu}}
\end{equation}
(in which $\varphi(z)$ is the chiral part of the scalar field\cite{byb})
which creates charge $1/\sqrt{\nu}$ in the charge sector (corresponding to charge 1 in physical units),
and a primary field $\sigma$ in the statistics sector: $a(z)=\sigma(z){\cal O}_{1/\sqrt{\nu}}(z)$.
(The field $\sigma$ has Abelian monodromy and is a simple
current, but these points will not be important for our discussion; the value of $\nu$ is determined so
that the correlator is single valued as one $z_i$ encircles another. In the Moore-Read example, $\sigma$
is the Majorana field.) Because the CFT enforces neutrality in the charge sector,
we have also inserted into the correlator $N/\alpha$ vertex operators $\mathcal{O}_{-\alpha/
\sqrt{\nu}}(w_j)$ of charge $-\alpha/\sqrt{\nu}$ at positions $w_j$; we refer to these as ``background
charges''. With the $w_j$ arranged on, say, a uniform grid with spacing $\propto \sqrt{\alpha}$, and
taking a limit $\alpha\rightarrow 0$, these negative charges create a continuous uniform background charge
distribution, and the function (using also a singular gauge transformation) represents a wavefunction
for particles in the lowest Landau level in a uniform magnetic field\cite{Moore1991,Read2009}. These
``conformal block'' trial wavefunctions include as examples the Laughlin and Moore-Read
states\cite{Moore1991}, and the Read-Rezayi series\cite{Read1999}. We note that, in line with Sec.\ 
\ref{sec:actionberry}, in this work we consider only one-component quantum Hall states; the particles 
carry no spin or other quantum numbers. There are straightforward generalizations to multicomponent 
conformal block states, to which our results can be extended; we return to this briefly at the end of Sec.\ 
\ref{sec:disc}.

Next we will begin in earnest to calculate the wavefunctions for a quantum Hall state on a surface in the
presence of the perturbed background metric specified by Eqs.\ (\ref{metric1}-\ref{metric3}), and also
a perturbed vector potential $\delta A$. Our final goal, as mentioned earlier, is to compute the Berry
curvature associated with perturbations of the Beltrami differential $\mu$. This requires that we calculate
the inner products of functional derivatives with respect to the Beltrami of the states. This can be
simplified if the states are normalized and depend holomorphically, or at least in an explicitly known
way, on the parameters that are varied adiabatically (see e.g.\ Ref.\ \onlinecite{Read2009}). Therefore,
we begin by examining the norms of the conformal block states in the presence of the background fields,
and for this we need the conformal blocks in the presence of the backgrounds. The latter are the subject of
the remainder of this section.

When we require the norms of the conformal block wavefunctions with respect to the usual inner product
that involves the integral over coordinates $z_i$ of the mod-square wavefunction, it is natural to invoke
a corresponding correlator of the non-chiral (i.e.\ left-right symmetric) CFT whose chiral parts were
used above; this is in fact the
starting point for the construction of chiral CFTs. Then in the most basic cases, this non-chiral
correlator can be factored into pieces respectively holomorphic (anti-holomorphic) in the particle
coordinates $z$. It would be natural to expect similar holomorphy properties in the presence of
perturbations of the background. However, when the almost-complex structure is perturbed by introducing
nonzero $\mu$, holomorphy in holomorphic coordinates like $F$ would be more natural than in the standard
coordinate $z$, and even then the wavefunctions are not expected to be holomorphic, but only up to the
non-holomorphic factor due to the background magnetic field (such as the familiar $e^{-|z|^2/4}$ in the
case of the plane) and (as we will see) the curvature of space
in the case of conformal primary fields with non-zero spin. In relation to the background fields $\mu$
($\bar{\mu}$) and $\delta \bar{A}$ ($\delta A$) (the relation of the latter with the physical potentials
$A$ is determined later in the paper, and the reason for the bar on $\delta A$ is conventional), one
would expect these conformal blocks to be holomorphic in their dependence on them also. However, it has
been known for a long time that factorization into a factor holomorphic in those fields times its conjugate
is not quite possible in general\cite{Quillen1985,Belavin1986,AG1986}, and
that this ``factorization anomaly'' is due to the anomalies in the gauge and coordinate-transformation
invariance
(or in current and stress correlations) in the chiral CFTs. To achieve factorization, and to ensure that
the non-chiral theory has no anomalies, there are additional factors multiplied into the blocks,
that are exponentials of integrals of local expressions (counterterms) that are neither holomorphic
nor antiholomorphic in the background fields\cite{AG1984,Polyakov1987,Polyakov1988}. These can be viewed
as being somewhat analogous to what occurs in the dependence on the particle coordinates due to the
background magnetic field.

Below, we will
first consider factorization in the presence of a U($1$) gauge field rather explicitly
in Subsection \ref{ssec:u1}. With the techniques demonstrated there, we will then move on to analyze the
gravitational case in Subsection \ref{ssec:grav}. Finally, in Subsection \ref{ssec:backcharge} we will
remark on the interplay between spatial curvature and the neutralizing background charges needed for
quantum Hall wavefunctions.


\subsection{Perturbed gauge field and gauge anomaly} \label{ssec:u1}

The simplest example of a holomorphic factorization anomaly is that associated with U($1$) gauge symmetry
in a chiral theory; we will discuss this case in detail. Consider a chiral conformal field theory with a
global U($1$) symmetry. There is a current $J$
(normalized so that
\begin{equation}
J=2\pi i\frac{\delta S}{\delta\, \delta\bar{A}}
\end{equation}
if $S$ is the action that depends on the corresponding vector potential $\delta A_\mu$) associated to
this symmetry by Noether's theorem. Generally, the current obeys the operator product expansion\cite{byb}
(or OPE; OPEs apply inside of expectation values, giving correlation functions)
\begin{equation}
J(z)J(0)\sim\frac{k}{z^2}+\ldots,\label{currentope}
\end{equation}
where $\ldots$ represent terms suppressed by positive integer powers of $z$, and the constant $k$ is called
the level of the current algebra. Similar relations hold for the anti-chiral
copy of the theory, with the substitutions $J\rightarrow\bar{J}$ and $z\rightarrow\bar{z}$. We will also
make use of the OPE
\begin{equation}
J(z)\phi(z')\sim \frac{q}{z-z'}\phi(z') +\ldots, \label{chargeope}
\end{equation}
as $z\to z'$ which holds within correlation functions. This means that $\phi_i$ is a primary
field\cite{byb} for the current
algebra as well as for the Virasoro algebra; $q$ is its U($1$) charge.

Let us now couple the chiral current to a gauge field $\delta\bar{A}$. In the chiral CFT, this has
the effect on (unnormalized) chiral correlators of inserting $e^{\frac{i}{2\pi}\int
d^2x\,\delta\bar{A}(w)J(w)}$
into any correlator. Then for the current expectation in the presence of the background gauge field we
have, using Eq.\ (\ref{currentope}),
\begin{align}
\bar{\partial}\langle J(z)\rangle&=\bar{\partial}\left\langle J(z)e^{\frac{i}{2\pi}\int
d^2x\,\delta\bar{A}(w)J(w)}\right\rangle_0, \nonumber \\
&\sim\frac{i}{2\pi}\int d^2x \delta\bar{A}(w)Z_{\delta\bar{A}}\bar{\partial}\langle J(z)J(w)\rangle_0,
\nonumber \\
&= -\frac{ik}{2}Z_{\delta\bar{A}}\partial\delta\bar{A},\label{u1anomaly}
\end{align}
where
\begin{equation}
Z_{\delta\bar{A}}=\left\langle e^{\frac{i}{2\pi}\int
d^2x\,\delta\bar{A}(w)J(w)}\right\rangle_0,
\end{equation}
and we have used the relation
\begin{equation}
\bar{\partial}\frac{1}{z}=\pi \delta(\mathbf{x}).
\end{equation}
This result is the gauge anomaly: the current $J$ is not holomorphic, or conserved (divergenceless),
as one might have expected naively, but instead $\bar{\partial}J=-i\frac{k}{2}\partial\delta\bar{A}$.
In addition, any charged primary fields in the chiral correlator act as $\delta$-function sources of
$\bar\partial J$.

Let us now consider what happens in the full non-chiral theory. In the left-moving part (if
the chiral or holomorphic part is right-moving), the OPEs have the same form, with the addition of bars
over $J$s, $z$s, $q$s and $k$. $\delta A$ is included by inserting $e^{-\frac{i}{2\pi}\int
d^2x\,\delta{A}(w)\bar{J}(w)}$, and then ${\partial}\bar{J}=
i\frac{\bar{k}}{2}\bar\partial\delta A$. Even when we set $\bar{k}=k$ (otherwise there is no possibility
for the non-chiral theory to be anomaly-free), the two pieces do not cancel in
$\bar{\partial}J-{\partial}\bar{J}$ (nor is the result gauge invariant). However,
if the logarithm of the partition function (unnormalized vacuum amplitude), or induced action, of the
non-chiral theory contains the additional local counterterm\cite{AG1984,Polyakov1988}
\begin{equation}
\delta S = \frac{k}{4\pi}\int d^2x\, \delta A\delta\bar{A}, \label{u1counter}
\end{equation}
which mixes the holomorphic and antiholomorphic components of the gauge field, then the net divergence of
the current is zero. We note that we can view this counterterm as producing an additional term proportional
to $\delta A_\mu$ in the expression for the current operator, and a delta-function term in the
current-current two-point correlator. Such terms are quite familiar in condensed matter physics
and include, for example, the diamagnetic term in the Kubo formula for electrical conductivity. Just like
in the non-chiral CFT, those terms are sometimes needed to satisfy the Ward identities which are
consequences of gauge invariance or of the divergencelessness of the current.

The counterterm has implications for the factorization properties of CFT correlators. The correlator of a
set of primary fields on a Riemann surface factorizes as
\begin{eqnarray}
\left\langle\prod_j
\phi_j(z_j,\bar{z}_j)\right\rangle&=&\exp\left(-\frac{k}{4\pi}\int d^2x\,
\delta A\delta\bar{A}\right) \\
&&{}\times\sum_{e,e'}N_{ee'}\Psi_e(\{z_j\}|\delta\bar{A})\bar{\Psi}_{e'}(\{\bar{z}_j\}|\delta A),\nonumber
\label{u1blocks}
\end{eqnarray}
where the conformal block functions $\Psi_e$ are independent of $\delta A$ (that is, the functional
derivative with respect to $\delta A$ vanishes) and are defined as
\begin{equation}
\Psi_e(\left\{z_j\right\}|\delta \bar{A})=\left\langle \exp\left(\frac{i}{2\pi}\int{d^2x\,
\delta\bar{A}(z)J(z)}\right)\prod_j\phi_j(z_j)\right\rangle_{\!\!0,e},
\end{equation}
and we denote by $\langle\dots\rangle_{0,e}$ the conformal block in the unperturbed
theory with $\delta A=0$, with $e$ and $e'$ as indices for a basis of conformal blocks in that system.
Also, we assume throughout that the left and right moving CFTs are the same, that the non-chiral theory
is ``diagonal'', and so are the non-chiral fields $\phi_j(z_j,\bar{z}_j)$ that we use, that is
$\bar{q}_j=q_j$. Then
$\bar{\Psi}_{e'}(\{\bar{z}_j\}|\delta A)=\overline{\Psi_e(\{z_j\}|\delta\bar{A})}$, and
$N_{e,e'}=\delta_{e,e'}$.

We now show how to obtain wavefunctions in the presence of a gauge field perturbation from factorization.
As we wish to postpone discussion of the background charges, and thus of true quantum Hall wavefunctions,
we will here continue to use a general correlator as we have so far. From Eq.\
(\ref{u1blocks}), crucially we include the square root of the non-holomorphic factor into the conformal
block to obtain what we can call a wavefunction $\psi_e$, in the presence of a background gauge field
$\delta A$:
\begin{equation}
\psi_e(\{z_j\}|\delta \bar{A},\delta A)=e^{-\frac{k}{8\pi}\int d^2x \delta\bar{A}\delta
A}\Psi_e(\{z_j\}|\delta\bar{A}).
\end{equation}

The results about the cancellation of the anomaly in the non-chiral correlator can now be rephrased in
terms of the behavior of these wavefunctions under a gauge transformation. As the expression is supposed
to give the wavefunction for any choice of $\delta A$, $\delta\bar{A}$, it should be correct to simply
substitute $\delta\bar{A}+\bar\partial\Lambda$ for $\delta\bar{A}$, and similarly for its conjugate.
($\Lambda$ is a real function of $z$, $\bar{z}$.) Using integration by parts, a similar calculation as
that above for $\bar\partial J$ shows that the wavefunction transforms by a phase that is one-half the
charge of each primary field times $\Lambda$ at the position of the field, plus an additional phase due
to the anomaly:
\begin{eqnarray}
\lefteqn{\psi_e(\{z_j\}|\delta \bar{A}+\bar{\partial}\Lambda,\delta A+\partial
\Lambda)=}&&\nonumber\\
&&\exp\left(-\frac{i}{2}\sum_j q_j\Lambda(z_j,\bar{z}_j)-\frac{ik}{16\pi}\int d^2x\, \Lambda
\delta B\right)\nonumber \\
&&{}\times\psi_e(\{z_j\}|\delta \bar{A},\delta A),
\end{eqnarray}
where $\delta B=\partial_x \delta A_y-\partial_y \delta A_x$
appears because of the anomaly. Note that, if the non-holomorphic term were absent,
the anomaly contribution would not be a phase factor; only a transformation by a phase factor can cancel
when the wavefunction is multiplied by its conjugate.

Finally, we check that the wavefunctions are holomorphic, in the sense that
their covariant derivative with respect to any of the $\bar{z}_j$'s vanishes. This will be useful in
ensuring that the conformal block indeed represents a lowest Landau level electron wavefunction,
and in correctly relating $\delta A$ in the CFT to perturbations of the physical vector
potential $A_\mu$. Taking the derivative and applying Eq.\ (\ref{chargeope}), we find for each $j$
\begin{equation}
\left(\partial_{\bar{z}_j}+\frac{i}{2}q_j\delta\bar{A}(z_j,\bar{z}_j)\right)\psi_e=0,\label{u1holo}
\end{equation}
provided $z_j$ is not equal to any $z_{j'}$, $j'\neq j$.

\subsection{Perturbed metric and gravitational anomaly} \label{ssec:grav}

When we consider perturbations of the background metric rather than in a U($1$) gauge field, the
relevant anomaly is the gravitational or coordinate-transformation anomaly. The analog of the current
is now the stress tensor $T$, which can be defined as \cite{Friedan1987}
\begin{equation}
T(z)=\pi\frac{\delta S}{\delta \mu}.
\end{equation}
and similarly for the antiholomorphic $\bar{T}$. The logic we used above to compute the gauge anomaly holds
essentially unchanged\cite{AG1984}, except that now the relevant OPEs are
\begin{align}
T(z)T(0) &\sim\frac{c}{2z^4}+\frac{2}{z^2}T(0)+\frac{1}{z}\partial T(0)+\ldots,\nonumber \\
T(z)\phi(0)&\sim \frac{h}{z^2}\phi(0)+\frac{1}{z}\partial\phi_i(0)+\ldots,
\label{stressOPE}
\end{align}
where $c=c_{\rm CFT}$ is the (right-moving) central charge of the chiral CFT, and $\phi_i$ is a primary
field with conformal weight $h_i$; for brevity, we will refer to $c_{\rm CFT}$ simply as $c$ throughout
this Section and the next, Sec.\ \ref{sec:berry}. Similar to the U($1$) case above, the {\it c}-number
term (with coefficient
$c$) in the $TT$ OPE implies the existence of an anomaly in purely chiral theories. As in the previous
case, when we combine the chiral and antichiral theories, we look for a counterterm to remove the
gravitational anomaly. There are two main complications here as compared to the U($1$) case. First,
the ``gauge'' group is not Abelian as it was for the U(1) gauge case, and so the local counterterms are
not simply bilinear in $\mu$, $\bar{\mu}$. Also, when the chiral and antichiral theories are combined
with the appropriate counterterm, there is still a relic of the anomaly remaining in the trace anomaly
associated with metric rescalings\cite{Friedan1987}.

H. Verlinde compiled a particularly concrete and general form for the nonholomorphic prefactors in a
correlator of primary fields in a non-chiral theory\cite{Verlinde1989}:
\begin{align}
\left<\prod_j
\phi_j(z_j,\bar{z}_j)\right>_{g}=&\exp\left(\frac{c}{12\pi}K[\mu,\bar{\mu},\Phi]-\sum_j\Phi(z_j)h_j\right)
\nonumber\\
&\times\sum_{e,e'}N_{ee'}\Psi_e(\{z_j\}|\mu)\bar{\Psi}_e'(\{\bar{z}_j\}|\bar{\mu}),\label{gravholfact}
\end{align}
where $h_j$ are the conformal weights of the primary fields $\phi_j$ (as before, we have $h_j=\bar{h}_j$).
Similarly, we assume the left-moving central charge $\bar{c}$ is equal to $c$, otherwise the non-chiral
theory would be anomalous. $\Psi_e$ is a conformal block given by
\begin{equation}
\Psi_e(\{z_j\}|\mu)=\left<\exp\left(-\frac{1}{\pi}\int
d^2x\,\mu(z,\bar{z})T(z)\right)\prod_j\phi_j\right>_{0,e}, \label{holblocks}
\end{equation}
We note that $\Psi_e$ is by definition independent of $\bar{\mu}$.
The functional $K[\mu,\bar{\mu},\Phi]$ is the Belavin-Knizhnik counterterm\cite{Belavin1986}
\begin{align}
K&=\int d^2x\,(1-|\mu|^2)^{-1}\left(\partial\mu\bar\partial\bar{\mu}-\frac{1}{2}\mu
(\bar\partial\bar{\mu})^2-\frac{1}{2}\bar{\mu}(\partial\mu)^2\right)\nonumber \\
&+\frac{1}{4}\int d^2x (1-|\mu|^2)\left(\frac{1}{2}e^\Phi
g^{\nu\lambda}\partial_\nu\Phi\partial_\lambda\Phi+\Phi R_{*}\right),
\end{align}
where $R_{*}$ is the Ricci scalar for the metric $g_{*}=e^{-\Phi}g$.
$K$ is the local counterterm needed to make the partition function invariant under coordinate
transformations, analogous to the U($1$) counterterm Eq.\ (\ref{u1counter}).

As above, we can define ``wavefunctions'' in the presence of a nontrivial metric as
\begin{align}
\psi_e(\{z_j\}|\mu,\bar{\mu},\Phi)=&\exp\left(\frac{c}{24\pi}K[\mu,\bar{\mu},\Phi]
-\sum_j\Phi(z_j)h_j/2\right)\nonumber\\
&\times\Psi_e(\{z_j\}|\mu). \label{gravwfs}
\end{align}
To cement the identification with electron wavefunctions, we check that the functions Eq.\
(\ref{gravwfs})
are annihilated by an antiholomorphic covariant derivative. We find, for each $j$,
\begin{equation}
\left.\left[\bar\partial-\mu\partial+h_j\left(-\partial\mu+\frac{1}{2}\bar\partial\Phi-\frac{\mu}{2}
\partial\Phi\right)\right]\right|_{z_j,\bar{z}_j}
\psi_e=0 \label{mulll}
\end{equation}
(the derivatives act on, and the functions are evaluated at, $z_j$, $\bar{z}_j$).
The terms multiplied by $ih_j$ agree with Eq.\ (\ref{spinconnF}), so this equation is equivalent to
\begin{equation}
\left(\partial_{\bar{F}}+ih_j\omega_{\bar{F}}\right)|_{F_j,\bar{F}_j}\psi_e=0.\label{gravholo}
\end{equation}

As in the gauge case, the nonholomorphic prefactors ensure that the wavefunction transforms under change of
coordinates only by a phase factor. In more detail, we first point out that, on calculating the expectation
of $\bar\partial T(z)$ in the conformal block, we obtain the anomaly
equation\cite{Polyakov1987,Polyakov1988}
\begin{equation}
\bar\partial T-\mu\partial T-2\partial \mu T = \frac{c}{12}\partial^3\mu
\end{equation}
(plus terms that appear when $z$ is at the location of one of the primary fields
$\phi_j$). Then including the counterterms, we find that under the
infinitesimal change of coordinates $z\rightarrow\zeta=z+f$ we have, through first order in $f$, $\mu$,
and $\Phi $,
\begin{align}
&\psi_e(\{\zeta_j\}|\mu^\zeta_{\bar{\zeta}},\bar{\mu}^{\bar{\zeta}}_\zeta,\Phi')=
\exp\left(\frac{ic}{12\pi}\int d^2x\, \mathrm{Im}(\partial
f\partial^2\mu)\vphantom{\sum_j}\right.\nonumber\\
&\quad\left.{}-i\sum_jh_j\mathrm{Im}(\partial f-\bar{\mu}\bar\partial f)\big|_{z_j,\bar{z}_j}\right)
\psi_e(\{z_j\}|\mu,\bar{\mu},\Phi),
\end{align}
where $\mu^\zeta_{\bar{\zeta}}$ and $\Phi'$ are $\mu$ and $\Phi$ in the $\zeta$ coordinates (see Sec.\
\ref{sec:prelim} for their form). Here ${\rm Im}\,(\partial f-\bar{\mu}\bar\partial f)$ is the local
rotation angle (where by a rotation we mean a transformation that leaves the local metric invariant).
These terms underline another way to view the non-holomorphic factors
$e^{-h_j\Phi/2}$ included in the wavefunctions, as follows. Usually, one considers conformal blocks
and uses holomorphic coordinates, in which the primary fields transform as tensors, with the conformal
weight as the (fractional) number of lower minus the number of upper indices; thus a field of weight
one transforms as the components of a one-form under a conformal
coordinate transformation\cite{byb}. To obtain the coordinate-invariant non-chiral correlators,
each lower $z$ index on the block must be contracted with a lower $\bar{z}$ index on the conjugate block
using the inverse metric tensor $g^{\mu\nu}$, which contains a factor $e^{-\Phi}$; we choose to absorb
the square root of each factor into the wavefunction. As the inverse vielbein $e^\nu_a$ is effectively
a square root of the inverse metric tensor, including the factors into the wavefunctions in
our case (in general, with $\mu\neq0$) can be viewed as transforming the blocks into tensors with only
internal indices, because the (inverse) vielbein converts the indices. The precise form corresponds to the
way that the factor $e^{\Phi/2}$ turns $dz+\mu d\bar{z}$ (which has an upper index) into the ``canonical''
vielbein one-form in Sec.\ \ref{sec:prelim}, in terms of which $ds^2$ is simply a modulus square, just
like the non-chiral correlators. Because of the use of the ``canonical'' vielbeins (and corresponding spin
connection), our wavefunctions transform by the local rotation (phase
factor) under coordinate transformation, which would not occur if they were genuine scalars. Genuine scalar
behavior (up to the anomaly contribution, anyway) can be restored by combining the coordinate
transformation with an internal rotation (i.e.\ by removing the one that is a consequence of the
``canonical'' vielbeins---see Sec.\ \ref{sec:prelim}).


In these expressions, we set $\delta A=\delta\bar{A}=0$; if we wish to turn on both perturbations,
we can simply include both sets of effects in a fairly straightforward way. The absence of
mixed gauge-gravitational anomalies in two dimensions ensures that the factorization formulas still
hold\cite{AG1984}. We note that then the conformal blocks depend on $\delta A_{\bar{F}}$ (a combination
of $\delta A$ and $\delta \bar{A}$) rather than on $\delta A$ only, just as they are functions of
particle coordinates $z_j$ and $\bar{z}_j$, not $z_j$ only; however, we will suppress both points in
our notation $\Psi_e(\{z_j\}|\delta\bar{A},\mu)$ for these blocks, to lighten notation. We note that in
the following Berry curvature calculations, we will in fact consider the two perturbations separately,
so that the explicit form of the combined expressions will not be required.

\subsection{Background charges, curvature, and magnetic field} \label{ssec:backcharge}

We now show what happens when background charges are introduced into the wavefunctions of the previous
subsections, so that we obtain true quantum Hall wavefunctions. As in the Moore-Read construction with
unperturbed background fields, the positions of these primary fields will be written as $w_k$, reserving
$z_j$ for particle coordinates. Using the wavefunctions from the previous subsections, we take a limit to
obtain wavefunctions
\begin{align}
&\psi_e(\{z_j\}|\delta \bar{A},\delta A,\mu,\bar{\mu},\Phi,\rho)=\nonumber\\
&\quad\exp\left(-\frac{1}{8\pi}\int d^2x \sqrt{\det g}\delta\bar{A}
\delta A+\frac{c}{24\pi}K[\mu,\bar{\mu},\Phi]\vphantom{\sum_j}\right.\nonumber\\
&\quad\left.{}-s\sum_{j'}\Phi(z_{j'})/2\right)
\lim_{\alpha
\rightarrow 0}\Psi_e(\{z_j\},\{w_k\}|\delta\bar{A},\mu).
\end{align}
Here we identified $h_j=s$ (the spin) for the primary fields $a$ that represent the particles, and put
$k=1$ to agree with the Moore-Read construction.
Also, $-\rho/\sqrt{\nu}$ is the density of background charge (in the $z$ coordinates and in the CFT units);
the limit is taken so that this
approaches a continuous density. We note that the part of the exponential factor containing $\Phi$
evaluated at the positions $w_k$ drops out, because the conformal weights $h_k$ for the background charges
are $O(\alpha^2)$. Taking the limit actually involves use of a singular gauge
transformation\cite{Moore1991}, as we explain next; this transformation is left implicit in this
expression.

Before the limit is taken, the wavefunction is not single valued as a $z_i$ makes a circuit (under
analytic continuation) around some $w_j$s\cite{Moore1991,Read2009}; without loss of generality, we can
think of a circuit that is not self-intersecting, contains no other $z_j$s, and is traversed in the
positive direction. The resulting phase factor is, in the limit,
\begin{equation}
\exp(i\theta_{c})=\exp\left(-2\pi i\nu^{-1}\int d^2x\,\,\rho\right),
\end{equation}
where the integral is over the interior of the curve, by using
Stokes's Theorem. The phase factor is removed from the wavefunctions by a singular gauge transformation,
whose form is clear before the limit (after the limit, it becomes very singular). The resulting
wavefunction, which is the one just defined, is single valued under analytic continuation, but due to the
singular gauge transformation there is an additional U(1) connection (vector potential) $A_\mu^0$
experienced by the particles, whose curl is the density (here we choose to use units where the particle
has charge $1$): $\partial_1 A^0_2-\partial_2 A^0_1=2\pi\rho/\nu$.

Combining the last connection with those due to $\delta A_\mu$ and $\omega_\mu$, as found earlier in
eqs.\ (\ref{u1holo}), (\ref{gravholo}), and comparing with the physical vector potential $A_\mu$ for
particles with charge $1$ (by convention) and zero orbital spin as in Eq.\ (\ref{vecholo}), we arrive at 
the identification of the space components of the vector potential, for our wavefunctions,
\begin{equation}
A_\mu=A_\mu^0-\frac{1}{2\sqrt{\nu}}\delta A_\mu-s\omega_\mu.\label{deltaAreln}
\end{equation}
The wavefunctions $\psi_e(\{z_j\}|\delta \bar{A},\delta A,\mu,\bar{\mu},\Phi,\rho)$ are LLL functions
for this $A_\mu$, as defined earlier in Sec.\ \ref{sec:prelim} (a similar result was found in Ref.\
\onlinecite{Klevtsov2013}).

Taking the curl of the last equation, we obtain for the densities (or two-forms)
\begin{equation}
B=\frac{2\pi}{\nu}\rho-\frac{1}{2\sqrt{\nu}}\delta B-s R,
\end{equation}
where the density $R=\partial_1\omega_2-\partial_2\omega_1$ is essentially the Riemann
tensor in two dimensions\cite{Bradlyn2014}. Rearranging, and setting $\delta B=0$, we have
\begin{equation}
\rho=\frac{\nu}{2\pi}(B+sR), \label{densityfield}
\end{equation}
This relation was recently derived using a different (but related) approach in
Refs.~\onlinecite{Klevtsov2014,Can2014,Can2014-1}. Anticipating the generalized screening arguments to be
made in the following section, in a wavefunction for a gapped quantum Hall state, $\rho(z,\bar{z})$ can
be identified with the local particle density (in units where the particles have charge $1$), up to possible
corrections higher order in gradients of the background fields. Integrating over the surface and using the
Gauss-Bonnet theorem then gives
\begin{equation}
N_\phi=\nu^{-1}N-s\chi
\end{equation}
where $2\pi N_\phi$ is the integral of $B$ ($N_\phi$ is the number of flux quanta), and $\chi=2-2{\cal G}$
is the Euler characteristic
($\chi=2$ for the sphere). This result contains the insight of Wen and Zee\cite{Wen1992}, that the shift
$\cal S$ in the number of flux quanta $N_\phi$ (${\cal S}=2s$ for the sphere) resembles the effect of an 
intrinsic spin $s$ carried by the particles, as spin couples to spatial curvature $R$.

\section{Berry curvature of trial wavefunctions} \label{sec:berry}

In this section, we will exploit the holomorphic factorization properties of conformal block trial
wavefunctions to compute Berry curvatures for quantum Hall states.
In Subsection \ref{ssec:cond} we consider as a warm-up exercise the response of a quantum Hall
system to a uniform perturbation in the vector potential (equivalent to using twisted boundary
conditions) on the torus with the standard metric, and from this derive the Hall conductivity. Using
similar reasoning, in Subsection \ref{ssec:centralcharge} we will show how the topological central charge
can be extracted as a Berry curvature.

\subsection{Hall conductivity} \label{ssec:cond}

Consider a quantum Hall system at filling factor $\nu$ on the torus. We take the metric of the torus to
be the standard one, $ds^2=|dz|^2$, and $z=x^1+ix^2$ with $x^1$, $x^2$ running from $0$ to $L$. As is
well known, addition of a constant vector potential $\delta A_\mu$ on the torus is equivalent (after a
gauge transformation) to twisting the boundary condition on the particles. We use the trial wavefunctions
$\psi_e(\{z_j\}|\delta \bar{A},\delta A,\rho)$ obtained above, in which we have dropped $\mu$, $\bar{\mu}$,
$\Phi$ as they are all zero. We take the background charge density, or $\rho$, to be uniform in these
coordinates:
$\rho=N/L^2$. For Laughlin states, the wavefunctions $\psi_e$ for $\delta A_\mu$ constant can be expressed
in terms of certain elliptic theta functions---see for example Ref.\ \onlinecite{Read1999}.

We now proceed to compute the Berry curvature associated with adiabatic transport in $\delta A_\mu$.
Here we invoke the hypothesis of generalized screening, that is, we suppose that in these wavefunctions
all correlations of local operators decay exponentially with distance. (Note that here we mean
quantum-mechanical expectations in the states with wavefunctions; in this section, angle brackets
$\langle\cdots \rangle$ always mean such expectations or inner products, defined by multiplying a
wavefunction and a complex conjugate wavefunction and integrating over particle coordinates with weight
$\sqrt{\det g}$ for each.) In the Laughlin states, generalized screening is simply screening in the plasma;
generalized screening and its consequences were discussed in Refs.\ \onlinecite{Nayak1996,Read2009}.
(Further, in Ref.\ \onlinecite{Bonderson2011}, mappings of some states onto actual plasmas were obtained.)
When the hypothesis holds, it implies\cite{Read2009} that the trial ground state wavefunctions, as we
have constructed them, are normalized {\em independently of $\delta A_\mu$} in the present case, at least
up to higher derivative terms that we can drop. (Locally, the wavefunctions are gauge equivalent to
those with $\delta A_\mu=0$, and so the normalization is
unchanged, as it is insensitive to global effects according to the hypothesis.) Then the Berry connection is
\begin{align}
\mathcal{A}_{\delta A}&=i\left\langle\psi_e\left|\frac{\partial\psi_e}{\partial \delta
A}\right.\right\rangle=-\frac{iL^2}{8\pi}\delta\bar{A},
\nonumber \\
\mathcal{A}_{\delta\bar{A}}&=i\left\langle\psi_e\left|\frac{\partial\psi_e}{\partial \delta
\bar{A}}\right.\right\rangle= -i\left\langle\left.\frac{\partial
\psi_e}{\partial\delta\bar{A}}\right|\psi_e\right\rangle=\frac{iL^2}{8\pi}\delta A,
\end{align}
where we have exploited the fact that the only dependence of the wavefunctions on
$\delta A$ (i.e.\ non-holomorphic in $\delta \bar{A}$) comes from the prefactor
$\exp\left(-\frac{1}{8\pi}L^2\delta\bar{A}\delta A\right)$. (The Berry
connection is really a matrix in $e$, $e'$, but off-diagonal elements are zero, and the diagonal elements
are independent of $e$, as shown.) Taking the exterior derivative (essentially, curl) with respect to
$\delta A$, we find for the components of the Berry curvature
\begin{align}
\mathcal{F}_{\delta A,\delta\bar{A}}&=-\mathcal{F}_{\delta \bar{A},\delta A}=iL^2\frac{1}{4\pi}.
\end{align}
Finally, using the correspondence Eq.\ (\ref{deltaAreln}), we have in terms of the physical vector potential
and Cartesian components
\begin{equation}
\mathcal{F}_{A_1,A_2}=-L^2\frac{\nu}{2\pi}.
\end{equation}
This agrees exactly with the Berry curvature in Eq.\ (\ref{u1berr}) for this perturbation, which was
obtained from the U($1$) Chern-Simons term in the induced action Eq.\ (\ref{Seff}). In fact, because
the vector potential
$A$ enters only through the combination $A_\mu+s\omega_\mu$, which here for
space components we can identify with $A_\mu^0-\frac{1}{2\sqrt{\nu}}\delta A_\mu$, this calculation (in
which only the space components of $A_\mu+s\omega_\mu$ are perturbed adiabatically) implies that the induced
action also contains the first and second Wen-Zee terms with coefficients as in that equation if we
identify $\overline{s}=s$, $\overline{s^2}=s^2$, similar to Ref.\ \onlinecite{Wen1992}. However, because
the change in metric can also enter in other ways, we cannot yet make the last of these identifications
rigorously.

Physically, we note that since a spatially uniform but temporally varying
vector potential on the torus corresponds to an electric field, this Berry curvature gives us
the Hall conductivity upon dividing by the area of the system, giving\cite{Niu1985} (with the correct sign)
\begin{equation}
\sigma^H=\frac{\nu}{2\pi}.
\end{equation}
As far as we are aware, this direct derivation of the Hall conductivity from conformal block
wavefunctions using the factorization behavior of the blocks, while immediate, has not previously appeared
in the literature.

We want to comment that, instead of constant $\delta A$, we could have considered functional
derivatives with respect to spatially-varying $\delta A$, as in Sec.\ \ref{sec:actionberry}. The derivation
would be similar, and
resemble more closely that in the following Subsection. Some care would be necessary in connection with
normalization and screening if this involved $\delta B\neq 0$ with $\rho$ fixed, but we will not enter
into this here.

\subsection{Central charge and Hall viscosity} \label{ssec:centralcharge}

Next, we consider the effect of metric perturbations in the wavefunctions $\psi_e(\{z_j\}|\delta
\bar{A},\delta A,\mu,\bar{\mu},\Phi,\rho)$. We work again on the torus, as in the previous Subsection;
we set $\delta A_\mu=0$; and we will hold the positions of the background charges fixed in the $z$
coordinates as the metric is varied, so $\rho=N/L^2$ is constant and fixed. In fact, this means we can
hold $A_\mu^0=A_\mu+s\omega_\mu$ fixed
as we vary the metric. This prevents us from picking up any Berry phase, like that in the previous
Subsection,
from the Hall conductivity. We consider adiabatic variation of the metric away from the standard flat one
$g_{\mu\nu}=\delta_{\mu\nu}$, viewing the Beltrami differentials $\mu$, $\bar{\mu}$ as the small
independent variations; they are always assumed to vary slowly on the scale of the particle spacing.
We further choose to fix $\det g=1$, which from Eq.\ (\ref{detgform}) implies that
\begin{equation}
e^\Phi=\frac{1}{1-|\mu|^2},
\end{equation}
so $\Phi$ is determined by $\mu$ also. We emphasize that all of this means that our variations do {\em not}
fix the magnetic field strength $B$; instead $B+sR$ is fixed, $R$ can be nonzero, and hence there can be
a change in $B$. These choices are simply the most convenient for our purpose.

Now we can again make use of the hypothesis of generalized screening. As the wavefunction transforms
as a scalar, up to phase factors, under a change of coordinates, we can go to holomorphic coordinates in
any local patch, and in these the conformal block has the same dependence on the coordinates $F$ as it
does in $z$ for $\mu=0$. (However, in these coordinates $\rho\propto e^\Phi=\sqrt{\det g}$ and these
are not independent of position in general; note that the wavefunctions include the effect of the
interactions between the background charges, and these depend on $\mu$ and $\rho$.) Then the norm-square of
the wavefunction is like a partition function of a two-dimensional system, which according to the
hypothesis of generalized screening is in a massive phase. As for the case of an ordinary non-chiral CFT
subjected to a perturbation that makes it massive (such as free fermions with a mass term, which also
arise below) and in a
perturbed background, the logarithm of the norm-square or partition function can be expanded as a series
of integrals of local expressions, in which each local expression is a coordinate-transformation
invariant constructed from the background fields and their derivatives, that is in the present case,
from the metric and the curvature density (two-form) $R$. Thus it has the form, in an arbitrary system of
coordinates $\zeta^\mu$,
\begin{equation}
\int d^2\zeta \left[a_0\sqrt{\det g} + a_1 R+\ldots\right].
\end{equation}
Here the constants $a_0$, $a_1$, \ldots  are scalars, independent of the $\mu$ that we used,
but dependent on the CFT used in constructing the state. (The fact that $a_0$ is independent of $\mu$ can
be seen for $\mu$ independent of position by using holomorphic coordinates and arguments from Section III
in Ref.\ \onlinecite{Read2011}, together with the $e^{-s\Phi/2}$ factors included in the wavefunctions
here. We emphasize the importance of the interactions among the background charges included in that
calculation; they are presumably also the reason why our result appears different from those in Refs.\
\onlinecite{Can2014-0,Klevtsov2014,Can2014,Can2014-1}, where those factors are not included in the
wavefunctions.)
The terms omitted are higher than second order in derivatives of $g_{\mu\nu}$; for example, one such term
contains $R^2$ (it can be more easily expressed as the square of the Ricci scalar, times $\sqrt{\det g}$).
Furthermore, $\int R$ is a topological invariant (the Gauss-Bonnet Theorem
again), and so does not change as $\mu$ is varied. Hence we conclude that the norm-squares
of our wavefunctions are independent of $\mu$ under the conditions we specified, through second order in
derivatives. Similarly, the wavefunctions for distinct $e$ are orthogonal. We can proceed to
the calculation of the Berry curvature similarly as in the case of the Hall conductivity above, and
of the Hall viscosity\cite{Read2009}. In fact, our approach reduces precisely to the approach of Ref.\
\onlinecite{Read2009} if we take $\mu$ (and hence $\Phi$) constant in space. In the wavefunctions,
the factors $e^{-s\sum_{j}\Phi(z_{j})/2}$, which are non-holomorphic in $\mu$ or
$\tau$, contain the same non-holomorphic factors $({\rm Im}\,\tau)^{Ns/2}$ that were crucial in the
calculation there.

Given that the wavefunctions are normalized up to a factor that is independent of $\mu$, we will see
in a moment that in order to compute the Berry curvature in the $\mu=0$ limit, we need only retain
nonholomorphic terms in the wavefunctions, and these only to order $|\mu|^2$. Expanding the counterterms,
these are
\begin{align}
&\psi_e(\{z_i\}|\mu,\bar{\mu},\Phi,\rho)\simeq\nonumber\\
&\quad \left(1+\frac{c}{24\pi}\int d^2z\,
\partial\mu\bar\partial\bar{\mu}\right)
\prod_i\left(1-\frac{s}{2}|\mu(z_i)|^2\right)\nonumber \\
&\quad{}\times\lim_{\alpha\to0}\Psi_e(\{z_i\},\{w_k\}|\mu),
\end{align}
and we recall that the conformal block $\Psi_e$ is holomorphic in $\mu$. (Here again we left the
singular gauge transformation implicit; readers who have concerns about the limit and gauge
transformation can simply postpone those operations to the end of the calculation.)

We define a Berry connection associated to general variations in $\mu$ in the usual way, though here
we do it using {\em functional} derivatives with $\mu$, $\bar{\mu}$, as
\begin{align}
\mathcal{A}^{ee'}_{\mu}(z,\bar{z})&=i\left<\psi_e\left|\frac{\delta}{\delta\mu(z,\bar{z})}\right|\psi_{e'}
\right>,
\label{A} \\
\mathcal{A}^{ee'}_{\bar{\mu}}(z,\bar{z})&=i\left<\psi_e\left|\frac{\delta}{\delta\bar{\mu}(z,\bar{z})}
\right|\psi_{e'}\right>.\label{Abar}
\end{align}
(Here it is important that the subscript $\mu$ or $\bar{\mu}$ means the function $\mu$, not an
ordinary index.) Using the orthonormality of the wavefunctions, we can see that the derivative
in Eq.\ (\ref{Abar}) is only sensitive to the nonholomorphic dependence of the chiral wavefunction on
$\bar{\mu}$. Similarly, to evaluate Eq.\ (\ref{A}) we can use the constancy of the norm of the wavefunction
to move the derivative with respect to $\mu$ onto the conjugate wavefunction. With these
observations, we find to first order that
\begin{align}
\mathcal{A}_{\mu}(z,\bar{z})&=\frac{i}{2}\rho s
\bar{\mu}(z,\bar{z})+\frac{ic}{24\pi}\partial\bar\partial\bar{\mu}(z,\bar{z}), \nonumber \\
\mathcal{A}_{\bar{\mu}}(z,\bar{z})&=-\frac{i}{2}\rho s\mu(z,\bar{z})-\frac{ic}{24\pi}\partial
\bar\partial\mu(z,\bar{z}).
\end{align}
In the first term of each expression, we used the fact that the expectation of the particle density
operator is $\rho$, independent of position.
(We dropped the indices $e$, $e'$, as once again $\cal A$ is diagonal in these indices, and independent
of $e$.)
Finally, upon taking a variational exterior derivative with respect to $\mu(z',\bar{z}')$ and its
conjugate, we arrive at the functional Berry curvature at $\mu=\bar{\mu}=0$:
\begin{equation}
\mathcal{F}_{\bar{\mu}\mu}=-\mathcal{F}_{\mu\bar{\mu}}=i\rho s\delta(z-z')+\frac{ic}{12\pi}
\partial\bar\partial\delta(z-z'). \label{mainresult}
\end{equation}
This is our first main result. The first term in the curvature is nonzero even for $\mu$ constant in space.
It is exactly the Berry curvature associated with adiabatic deformations of the modular parameter $\tau$
of the torus as computed in Ref.\ \onlinecite{Read2009}, which gives the Hall viscosity\cite{Avron1995}.
The second term contains the central charge $c$ of the (chiral) CFT.

To compare with the induced action in Eq.\ (\ref{Seff}), we first note that we have fixed the space
components of $A_\mu^0$,
so by design the contribution of the U($1$) Chern-Simons and Wen-Zee terms of the form found in the
Hall conductivity calculation above is absent. However, uniquely, the first Wen-Zee term also contributes
to Berry curvature through terms in the action of quadratic order in $\mu$ that arise from the {\em time}
component $\omega_0$ of the spin connection. To the order needed, we have for $\omega$ in terms of $\mu$
\begin{align}
\omega_z&= -i\bar\partial\bar{\mu}+O(\mu^2),\label{omegapert1}\\
\omega_{\bar{z}}&=i\partial\mu+O(\mu^2),\label{omegapert2}\\
\omega_0&=\mathrm{Im}\left(\mu\partial_0\bar{\mu}\right)+O(\mu^3).\label{omegapert3}
\end{align}
Inserting these expressions in the induced action, we find that the Berry curvature above corresponds
with that obtained from the Chern-Simons terms under the conditions as stated, with the Hall viscosity that
agrees with the coefficient in the first Wen-Zee term\cite{Read2009b,Hoyos2012}, while if we assume that 
$\overline{s^2}=s^2$ then the topological
central charge in the gravitational Chern-Simons term is exactly the chiral central charge in the CFT
underlying the trial wavefunctions, as expected. This is our second main result; it is discussed further 
in the following Section \ref{sec:disc}.

We want to point out that the calculation in this subsection also applies to conformal-block trial
wavefunctions for particle systems without a background magnetic field; such wavefunctions can even be
used for systems of anyons. In this case, there are no complications related to a neutralizing
background charge density. One example is a $p-ip$ paired state
of fermions, represented by a conformal
block in a free chiral Majorana fermion theory; it corresponds to the Moore-Read state without the
charge sector. (In this example the norm-square of the wavefunction is precisely the partition function
of a massive Majorana field in a background metric, which we used above to illustrate the case with
background charges.) Then in all such examples the induced action lacks the U($1$) gauge Chern-Simons and
both Wen-Zee terms, because the filling factor is infinity and cannot occur. For the $p-ip$ example, by
a calculation as here, the gravitational Chern-Simons term occurs
with coefficient $c=1/2$, while the Hall viscosity (the value of which contains the mean particle
density in place of $\rho$, and $s=1/2$) must now be accounted for through an
``Euler current'' term in the induced action\cite{Golkar2014}.

\section{Discussion}\label{sec:disc}

In this Section, we comment on our results and their applications, and discuss some of the remaining
issues. We concentrate on the one-component quantum Hall examples as before, but discuss multicomponent 
generalizations at the end.

\subsection{General discussion for one-component systems}

First, we emphasize again that we have made the general point that the Chern-Simons terms (the terms that
are not locally invariant) in the induced action can be related to Berry curvature calculations for the
ground state in the presence of perturbed background fields. For non-relativistic systems, we described
the geometric aspects for metric perturbations, and how the LLL condition can be formulated. We showed that
for trial wavefunctions obtained from conformal blocks, the topological central charge $c$ is equal to the
chiral central charge $c_{\rm CFT}$ of the CFT, up to one assumption we discuss below; this validates our
procedure. But the initial point and the set-up
is much more general. It can be implemented numerically, if the ground state wavefunction in the presence
of (small) perturbations in the backgrounds can be found: in particular, of the metric (or vielbein),
to obtain the central charge. The most straightforward way to do so is to take $\mu=\mu_0
\exp(ik_\nu x^\nu)$, with $\mu_0$ constant, and the wavevector $\bf k$ compatible with the
geometry of the torus, in which case we see that there will be a Berry curvature which for the
conformal-block trial states takes the form
\begin{equation}
\mathcal{F}_{\bar{\mu}_0,\mu_0}=i\left(\rho s -\frac{c|\mathbf{k}|^2}{48\pi}\right) L^2
\end{equation}
in $\mu_0$ space (when holding $A_\mu+s\omega_\mu$ fixed), where 
$c=c_{\rm CFT}$. We mention that this expression allows us to see immediately (by comparing with
Refs.\ \onlinecite{Avron1995,Read2009}) that the topological
central charge contributes to the $O(|\bf k|^2)$ part of the Hall viscosity\cite{Abanov2014-0,Can2014},
under the conditions we used to calculate it.
Similarly, one could work on the sphere, using spherical harmonics.
Further, the method is not restricted to the topological phases considered here,
that can be modeled by wavefunctions obtained from conformal blocks. The approach also allows considerable
flexibility. For example, it is not necessary to vary the magnetic field $B$ so that $B+sR$ remains fixed:
one could work at fixed $B$ (or $A_\mu$) instead.

There is however one important point about what we have said. First we remark that in the general case 
we have written the coefficient of the second Wen-Zee term as $\overline{s^2}$, which is natural if it 
is viewed as a Hall conductivity for orbital spin, and not all particles have the same spin. Likewise, 
the coefficient of
the first Wen-Zee term is written in terms of $\overline{s}$, which is also related to the shift on the
sphere\cite{Wen1992}, ${\cal S}=2\overline{s}$. Now as discussed in Sec.\ \ref{sec:actionberry}, from
a bulk point of view the second Wen-Zee and the gravitational Chern-Simons terms cannot be distinguished.
Then one can expect to obtain only the apparent central charge $c-12\nu\overline{s^2}$ from the bulk induced
action or Berry curvature, after the coefficients of the U($1$) Chern-Simons and the first
Wen-Zee term have been determined from the Hall conductivity and Hall viscosity (which we know how to
obtain from Berry curvature calculations that do not involve spatial variation of $\mu$). If we use the
same technique as in the preceding calculation, of varying the metric holding
$A_\mu+\overline{s}\omega_\mu$ fixed, then in general the Berry curvature contains the coefficient
\begin{equation}
c-12\nu\left(\overline{s^2}-\overline{s}^2\right)
\end{equation}
in place of $c$. (The combination $\overline{s^2}-\overline{s}^2={\rm var}\, s$ can be called the variance
of the orbital spin.) Likewise, when working with $A_\mu$ fixed, the coefficient will be
$c_{\rm app}=c-12\nu\overline{s^2}$. In general, if we assume that the coefficients of the distinct terms are
as written in Sec.\
\ref{sec:actionberry}, then in a numerical study we will need additional information to determine
${\rm var}\, s$ separately from $c$, where these are defined using the coefficients
of terms that can only be distinguished using information from the edge of the system. Hence an
additional bulk Berry curvature calculation cannot resolve this issue, even in principle.

In the context of our calculation for trial states, strictly speaking we find only that $c-12\nu{\rm var}\,
s$ is equal to $c_{\rm
CFT}$, the central charge of the chiral CFT used in the construction of the states. If we make the natural
assumption that for these states $\overline{s^2}=s^2$ (as well as using $\overline{s}=s$---see Ref.\
\onlinecite{Read2009}), then $c=c_{\rm CFT}$, as stated above.

A more general argument is also available. The second Wen-Zee term and
the gravitational Chern-Simons term, while equivalent in the bulk, do differ at the edge of the system.
The expectation is that the topological central charge $c$ multiplies the gravitational Chern-Simons terms
(as we have written in Sec.\ \ref{sec:actionberry}).
It then describes the inflow of energy and momentum onto the edge under perturbations of the vielbeins, an
effect that cancels the gravitational anomaly (i.e.\ in energy and momentum conservation) in the edge
theory\cite{Callan1985} (see also Ref.\ \onlinecite{Stone2012}). On the other hand, the Wen-Zee terms
contain the spin connection $\omega_\mu$, and the corresponding gauge invariance under internal rotations
is used to account for rotation invariance in ambient space (in a local time slice). The latter symmetry
is obviously lost at an edge, and in the edge theory there is in general no Lorentz invariance that can
take its place, even at low energies. Hence for the second-Wen-Zee term, in particular, it is not at all
clear that there should be any well-defined corresponding anomaly in the boundary effective theory.
[We should point out that if a system has full relativistic Lorentz invariance, both in the bulk and the
edge, then the gravitational and internal Lorentz-symmetry anomalies are essentially
interchangeable\cite{Bardeen1984} (but even in this case, the Lorentz invariance on the boundary is in
one dimension fewer than in the bulk). We cannot expect the same to hold in the non-relativistic situation
considered here.] Then we can make the field-theoretical argument that the coefficient of the gravitational
Chern-Simons term {\em has} to be the topological central charge $c$ of the edge theory (just as written
in Sec.\ \ref{sec:actionberry}). For the
conformal-block trial states, the topological central charge of the edge theory is supposed to be the same
as the central charge of the CFT used in the construction (this is part of a conjecture that goes back to
Moore and Read\cite{Moore1991}, and corresponds to a result of Witten\cite{Witten1989}), and so our 
conclusion from the calculation is that
\begin{equation}
{\rm var}\,s=0.
\end{equation}
In general, this argument shifts the discussion from one about the topological central charge to one about
the value of $\overline{s^2}$: $\overline{s^2}$ can be obtained from $c_{\rm app}$ if the value of $c$ is 
taken
as known. We remark that our result ${\rm var}\, s=0$ for the trial states considered here disagrees
in a number of cases with the recent results of Ref.\ \onlinecite{Gromov2014} (in which $c=c_{\rm CFT}$ 
was also obtained); after the appearance of
the first version of the present paper, these authors discovered an error in their calculation, 
and their revised result\cite{GromovErratum} agrees with ours.

It may be thought that for Abelian quantum Hall states, the value of $\overline{s^2}$, as well as that
of $\overline{s}$, or the shift, can be predicted analytically from the work of Wen and Zee
(WZ)\cite{Wen1992}. In the case of the shift, its value is easily obtained, at least in many cases, from
other approaches such as trial wavefunctions, but the situation is less clear for $\overline{s^2}$. 
For conformal-block states, $\overline{s^2}=\overline{s}^2=s^2$, where $s$ is the conformal weight of 
the field $a$ that represents the particles, seems to be natural as we have said, and is supported by 
our results under an assumption. In particular, for the Laughlin state at filling $\nu=1/Q$, we agree with 
WZ that $\overline{s^2}=Q^2/4$. In certain other cases, obtaining it is straightforward. 
In the case of non-interacting fermions filling (integer) $\nu$ Landau levels, the orbital spin is ${\cal 
N}+1/2$ for the $\cal N$th Landau level (${\cal N}=0$, $1$, \ldots; note that the angular momentum is 
minus this spin in our conventions), which agrees with the angular momentum due to the cyclotron motion. 
In this case $\overline{s^2}$ is calculable, and the result should
be correct (it seems to have motivated WZ\cite{Wen1992}). Some further examples are discussed at the end 
of this section.

For general Abelian states, WZ used an approach in which there are in effect several types of particles,
with different charges under the U($1$) gauge field. (These types of particles may correspond to the use of
hierarchical constructions, or other techniques.) They argued that by analogy with the charge,
these particles can have different spins, through which they couple to spatial curvature. They begin with
an Abelian Chern-Simons effective (not induced) action (which unlike the induced action we use in this
paper) describes the dynamics of the low-energy states of the systems, which are not yet integrated out. It
contains as its dynamical variables one or more gauge fields $a_\mu$, and the coupling to the background
gauge field $A_\mu$ (spin connection $\omega_\mu$) is through a charge vector $t$ (spin vector $s$) that
has components in the index space of the fields $a_\mu$ (which we suppress in our notation). On
integrating out the dynamical gauge fields $a_\mu$, they obtained an induced action of the same form as
our $S_{\rm eff}$, in which $\nu$, $\overline{s}$, and $\overline{s^2}$ are given by expressions containing
the $K$-matrix of the Abelian state and the charge and spin vectors $t$ and $s$. (However, they omitted the
gravitational Chern-Simons term entirely, which should arise from the derivation; this has been corrected
recently\cite{Gromov2014}.) Their main result was explaining the shift $\cal S$ in terms of
$\overline{s}$, as we have already mentioned.

However, WZ also obtained expressions for the spin of the quasiparticle excitations of the Abelian states
(as well as expressions for the charge and statistics that are not in question). They pointed out that,
with their expressions, the spin-statistics theorem does not hold (this is reaffirmed in their
Erratum\cite{WenZeeErratum}).
In two dimensions, for quasiparticles with Abelian statistics, the spin-statistics theorem says that
\begin{equation}
\theta =2\pi s \; ({\rm mod}\, 2\pi),
\end{equation}
where $\theta$, which is only defined modulo $2\pi$, is the Berry phase picked up on adiabatically
exchanging two
identical such quasiparticles along a counter-clockwise semicircle (and removing the Aharonov-Bohm
phase\cite{Arovas1984}), and $s$ is the spin of one of the
quasiparticles, under the same sign convention as in this paper. For a simple topological derivation
that holds both relativistically and non-relativistically, see Ref.\ \onlinecite{Wilczek1984}.
In Abelian Chern-Simons gauge theories, as well as in
related CFTs, the statistics of the quasiparticles, as well as their spins, are quadratic expressions
in their quantum numbers (the fluxes in the dynamical gauge fields). However, while for the statistics
WZ agree that the expression is quadratic, WZ's spins contain a term linear in the quantum numbers, that
is also linear in the spin vector; this is the source of the disagreement. It is also a requirement
that if the statistics of a quasiparticle of type $\alpha$ is $\theta_\alpha$,
then the statistics of its antiparticle type $\bar{\alpha}$ (which has the opposite sign quantum numbers)
is $\theta_{\bar{\alpha}}=\theta_\alpha$ (all of these are modulo $2\pi$). If the expression for the spin
contains a term linear in the quantum numbers (and in the spin vector), then it is not possible to satisfy
this relation either. We note that, in the modular tensor category point of view, which captures the
strictly topological properties of a topological phase, spin is only defined modulo integers, just
as statistics is only defined modulo $2\pi$ for Abelian quasiparticles. (These statements were reviewed
in e.g.\ Ref.\ \onlinecite{Read2009}.) This makes sense, because it is always possible in principle to
make some local, bosonic, excitation near a quasiparticle that changes its angular momentum by an integer,
without changing anything topologically. For quasiholes in conformal-block states on the sphere, the
spins were shown to agree with the conformal weights\cite{Read2008}, and not only modulo integers,
and to obey the spin-statistics theorem. The Laughlin state is one of these, and the result for that is in
direct disagreement with WZ.

The exception to the difficulty in satisfying the requirements in the presence of a linear term in the spin
is that, in some cases, the quantities can be equal modulo $2\pi$,
even though they are not equal as real numbers. This is what occurs for the case of $\nu$ filled Landau
levels. There, the spin of a hole in one of the filled Landau levels is $1/2$ (modulo an integer), and
spin-statistics is satisfied. The question of antiparticles does not arise, because additional particles
cannot be created in the filled Landau levels (other than to destroy a hole) to make the antiparticle of
the hole. Particles can, however, be created in one of the empty Landau levels above, and are again
fermions with half-integer spin. It is not clear if they are described by the $\nu$-component dynamical
Chern-Simons gauge theory, however.

In view of these observations, we do not accept in general the spin values assigned to quasiparticles by
WZ's theory. This may also cast some doubt on the values of $\overline{s^2}$ that were obtained by WZ; 
at the least, aspects of the derivation they gave (some of which was also used in the recent Refs.\ 
\onlinecite{Abanov2014,Gromov2014}) seem questionable.

\subsection{Multicomponent systems}

Finally, we return briefly to the topic of extending the results for conformal-block trial states to 
multicomponent conformal-block states. This can mean either that the underlying particles, whose 
wavefunction we are describing, are identical, but carry
additional quantum numbers or ``internal'' indices, which can represent internal (not orbital) spin or 
layer indices, and so on, or that there are simply different types of particles present, so that particles 
of distinct types can be distinguished from one another. (In the remainder of this section, we use indices 
$a$, $b$, \ldots $=1$, \ldots, $\cal N$ to represent particle types, and emphasize that these do not in 
general correspond to
quasiparticle types which were mentioned just above.) In addition, there can be background magnetic
fields, which may be different for different types of particles, and not all derived from a single U($1$) 
magnetic field as was the case previously. In the conformal-block trial wavefunctions, each particle of 
type $a$ has an orbital spin $s_a$, equal to the conformal weight of the primary field that represents 
that particle type in the CFT; clearly, particle types that are related by a symmetry will have the same 
$s_a$. We begin with the Berry curvature for varying the metric, with the background charge densities 
held fixed, and additional vector potentials (corresponding to $\delta A_\mu$ in Sec.\ 
\ref{sec:berry}; both the magnetic fields and the vector potentials are discussed further below) set to 
zero. Then it is almost immediate that the Berry curvature has the same form as obtained in the 
one-component cases in Sec.\ \ref{ssec:centralcharge}, with the Hall viscosity equal to one
half the density of orbital spin (i.e.\ to $\frac{1}{2}\sum_a n_a s_a$, where $n_a$ is the number density 
for each type $a$), and $c=c_{\rm CFT}$ is the central charge of the CFT used in the construction. In 
this sense, the preceding result generalizes straightforwardly.

For the remainder of the induced action, or of the responses to background fields, the structure is more 
involved, and here a brief sketch will have to suffice. In general, the presence of more than one particle
type will imply that there are additional conserved quantities in the system, such as the number of 
particles of each type or for each index value (these correspond to U($1$) symmetries), and there could 
be further symmetries that map one index value to another [so that some U($1$)s are in fact part of a 
non-Abelian group, such as SU($2$)], as in the case of internal (not orbital) spin. If the largest possible 
group of continuous ``internal'' Lie-group symmetries is identified, then we will assume that it has the 
form of a direct product of one or more U($1$) factors with one or more simple Lie groups, each of 
which might be SU($n$), Spin($n$), Sp($2n$) (each for some value of $n$), or one of the 
exceptional simple Lie groups, and possibly a quotient by some discrete subgroup must also be taken, 
to precisely describe the group. (We ignore the latter, as we usually focus on the Lie algebra level 
of description.) We have mentioned that we can include background magnetic fields; we assume 
that these take the form of U($1$) field strengths only. As we intend
the symmetry group we identified to be a symmetry of both the CFT used and of the states, the background 
magnetic fields must respect the symmetries; if one begins with a non-Abelian group and introduces 
a background field strength this breaks the symmetry down to some U($1$)s by assumption, and it is only 
the remaining unbroken symmetry to which we will refer as the ``symmetry''. 

The induced action should now be extended to include vector potentials corresponding to each of these 
continuous symmetries. Apart from the gravitational Chern-Simons term and second Wen-Zee term, whose form 
is unchanged from the one-component case, the Chern-Simons terms allowed in the induced action are Abelian 
and non-Abelian Chern-Simons terms that involve only the gauge (vector) potentials, and the first Wen-Zee 
term, which involves $\omega_\mu$ in addition to a fixed linear combination of the U($1$) vector 
potentials. The U($1$) potentials will be important to us, so (choosing a particular basis) we will denote
them by $A_\mu^{(a)}$, one for each particle type $a$. 

Next we consider the conformal-block trial wavefunctions obtained by a Moore-Read construction subject to 
the assumptions so far mentioned. In general, we may introduce into the construction both the background 
spatial metric and corresponding spin connection, and for each continuous symmetry (whether Abelian or 
non-Abelian) {\em for which there is a conserved current in the CFT}, a corresponding 
vector potential (generalizing $\delta A_\mu$ in Sec.\ \ref{sec:berry}). We note that there may not be such 
currents in the CFT for every conserved quantity (or symmetry) in the system; for example, in 
the Majorana field theory addressed at the end of Sec.\ \ref{sec:berry}, there is a conserved particle 
type, but no U($1$) current. But for each current operator in the CFT, there are non-zero gauge 
anomalies, and these produce corresponding Berry curvatures (obtained with 
the remaining vector potentials and metric held fixed). For the U($1$) vector potentials, these can 
take the form of a matrix of terms, as we will describe in a moment. It is only 
for these U($1$)s that background magnetic fields can be present in the state, because such a field has 
to be simulated by background charge densities as in the one-component case addressed so far. 

We must also relate the ``physical'' vector potentials to the corresponding ones used in the conformal 
blocks. We focus on the U($1$) vector potentials; the non-Abelian ones can be handled in a similar way, but 
with fewer complications. For the U($1$) vector potentials, we first make a further assumption about what 
our trial states describe, for definiteness: namely, we assume that the underlying particles all have 
zero orbital spin. (This was the assumption above, also.) We will label the U($1$) vector potentials 
$\delta A_\mu^{(u)}$ (corresponding to U($1$) current operators in the CFT) by indices $u$, $v$, \ldots 
$=1$, \ldots, $\cal M$ (${\cal M}\leq {\cal N}$). When we examine the holomorphy condition obeyed by the 
wavefunction in each particle coordinate, similar to Sec.\ \ref{ssec:backcharge} it contains the 
background magnetic field seen by
that particle type, the spin connection times $s_a$ for that type, and some combination of the background 
gauge field perturbations consisting of the U($1$) potential $\delta A_\mu^{(u)}$ (and possibly also 
some non-Abelian vector potentials), and this combination must be identified with the ``physical'' 
potential, which consists of $A_\mu^{(a)}$ and some non-Abelian potentials. Rearranging slightly, for the 
U($1$) parts this takes the form
\begin{equation}
A_\mu^{(a)}+s_a\omega_\mu=P_{au}\delta A_\mu^{(u)},
\label{constraints}
\end{equation}
where $P$ is an ${\cal N}\times{\cal M}$ matrix, and we omit the vector potentials $A_\mu^{0u}$ 
representing background magnetic fields, as they do not affect the argument.  
This relation implies that the shift in the number of flux seen by particles
of type $a$ is always $2s_a$, even when there is no extensive background magnetic field (for the $p-ip$
example, this was pointed out in Ref.\ \onlinecite{Read2000}.) We can assume that $P$ is an injective map 
(it has rank $\cal M$), because otherwise some of the U($1$) currents and vector potentials can be 
eliminated by a change of basis, reducing $\cal M$. 

The subsequent analysis is most conveniently handled in distinct cases. First we describe what we will 
call the non-degenerate case. In this case, $P$ is invertible, so in particular ${\cal N}={\cal M}$, and 
a similar statement holds for the non-Abelian gauge fields, if any. In this case we obtain states that 
describe true multicomponent quantum Hall-like states that have a gap in the bulk excitation spectrum 
for all excitations, so that the induced action is local. As $P$ is invertible, for the U($1$) gauge 
fields we can choose the basis for the space with indices $u$ such that $P$ is the identity, and then 
use only $a$, $b$ indices. Then the part of the induced action containing the U($1$) vector potentials 
takes the form
\begin{equation}
\frac{1}{4\pi}\int d^3x\,\widehat{\epsilon}^{\mu\nu\lambda}\sum_{a,b}K^{-1}_{ab}
(A_\mu^{(a)}+s_a\omega_\mu)\partial_\nu(A_\lambda^{(b)}+s_b\omega_\lambda).
\end{equation}
where $K^{-1}_{ab}$ is the matrix of gauge anomalies in the U($1$) currents (suitably normalized), 
or Hall conductivities, with the $a$, $b$ entry corresponding to currents of types $a$, $b$ in the CFT; 
it is real, symmetric, and positive definite, because the CFT used is chiral and should be 
unitary\cite{Read2009}. This form resembles the induced action in WZ, but has additional gauge
potentials. (We point out that the WZ effective action\cite{Wen1992} actually has additional U($1$) 
symmetries, so that such background gauge potentials could have been introduced there also.)
The remaining Chern-Simons terms in the induced action are the gravitational 
Chern-Simons term, which was already discussed, and the non-Abelian (pure gauge) Chern-Simons terms. 
Thus we have determined the coefficients of all Chern-Simons--type terms in the induced action, using 
Berry curvature. We see that the analog of $\nu\overline{s^2}$ is
\begin{equation}
\nu\overline{s^2}\longrightarrow\sum_{a,b}K^{-1}_{ab}s_a s_b,
\end{equation}
just as in WZ\cite{Wen1992}, however in the present very general point of view, it is not clear if 
there is always
a unique way to identify $\nu$ and so separate it from this expression. (There are different types of
particles, and different magnetic fields for each, and thus different filling factors; indeed the filling
factor has become a matrix.) Then if one calculates the
Berry curvature for varying the background metric with all the vector potentials including 
$A_\mu^{(a)}$ (instead of $A_\mu^{(a)}+s_a\omega_\mu$) for all $a$ held fixed, the coefficient will be
$c_{\rm app}=c-12\sum_{a,b}K^{-1}_{ab}s_a s_b$ (instead of $c$). In these more general examples, we cannot
necessarily obtain a simple relation between some $\bar{s}$ and $\overline{s^2}$. On the other hand,
we have seen that an analog of ${\rm var}\,s=0$ does hold in these states, in the sense that
the Berry curvature with $A_\mu^{(a)}+s_a\omega_\mu$ (for all $a$) and non-Abelian vector potentials 
held fixed contains just the coefficient $c$, with no correction like ${\rm var}\,s$.

In the degenerate case (we assume it is the U($1$) sector that causes the degeneracy; again, the 
non-Abelian analog is similar), the trial wavefunctions are expected to possess long-range correlations 
in some off-diagonal local operators, and so for a Hamiltonian that is short-range and respects the 
symmetries they describe
gapless phases of matter that exhibit broken symmetries; particular 
examples include quantum Hall ferromagnets, the $p-ip$ paired superfluid as described
in Sec.\ \ref{ssec:centralcharge}, anyon superfluids, and combinations of these. Accordingly, the 
full induced action may not be local. We can nonetheless analyze the Berry curvatures possessed by the 
trial states, and we will describe these in terms of Chern-Simons terms as before; these terms are likely 
to appear in any induced action for the system. Now the relations (\ref{constraints}) act as constraints 
on some combinations of $A_\mu^{(a)}$, which cannot be varied freely at fixed $\omega_\mu$ (by varying 
$\delta A_\mu^{(u)}$) as they could in the non-degenerate case. $P$ can be 
inverted only for those $A_\mu^{(a)}+s_a\omega_\mu$ that satisfy these 
constraints. For such background gauge fields, the U($1$) part of the induced action takes the form
\begin{align}
\frac{1}{4\pi}\int d^3x\,\widehat{\epsilon}^{\mu\nu\lambda}\sum_{u,v}K^{-1}_{uv}&
[P^{-1}(A_\mu+s\omega_\mu)]_u\nonumber\\
&{}\times\partial_\nu[P^{-1}(A_\lambda+s\omega_\lambda)]_v.
\end{align}
Here $K^{-1}_{uv}$ is a real symmetric positive-definite matrix of the gauge anomalies as before, and
$[P^{-1}(A_\mu+s\omega_\mu)]_u$ is the inverse image of the set of $A_\mu^{(a)}+s_a\omega_\mu$, which 
exists because of the constraint, and is unique because $P$ is injective. We have already described the 
Berry curvature for varying the metric with $\delta A_\mu^{(u)}$ fixed for all $a$ (which gives the Hall 
viscosity and $c$), and for varying the $\delta
A_\mu^{(u)}$ with fixed metric (which gives the matrix $K_{uv}^{-1}$). (We should also point out that 
in these cases the Hall viscosity and the shifts are not fully accounted for by these terms in the induced 
action; an example of this was already seen at the end of Sec.\ \ref{ssec:centralcharge}.) It remains 
to discuss the analog of $\nu\overline{s^2}$. Previously, this could be obtained by varying the metric with 
$A_\mu^{(a)}$ held fixed, at least by using the induced action, and probably also as a Berry curvature.
In the present case, the same can be done when the spin vector with components
$s_a$ lies in the image space of $P$ (i.e.\ it is a linear combination of the columns of $P$). Then it
is possible to hold $A_\mu^{(a)}$ fixed as the metric is varied, and still satisfy the constraint. In
this special case, the analog of $\nu\overline{s^2}$ is $\sum_{u,v}K^{-1}_{uv}(P^{-1}s)_u(P^{-1}s)_v$.
(This case subsumes the non-degenerate cases as well.) But in the general case, in which the spin vector 
$s_a$ does not lie in the subspace, when $\omega_\mu$ varies (because of the
variation in metric), $A_\mu^{(a)}$ must change somewhat, so as to remain in the constrained 
subspace at each point in space. While varying with $\delta A_\mu^{(u)}$ held fixed makes sense, keeping 
$A_\mu^{(a)}$ fixed except so as to satisfy the constraint requires a projection of the $A_\mu^{(a)}$ 
before the variation into the constrained space which has been translated because of the change in 
$\omega_\mu$. In the absence of an inner product on the space with indices $a$, there is no unique way 
to do this, or to define $\nu\overline{s^2}$; the answer depends on exactly 
what is held fixed, and there is no preferred choice. 

For cases in which degeneracy occurs in the non-Abelian sector, a different phenomenon can occur, in which 
the symmetries generated by the currents in the CFT form a proper subgroup of the non-Abelian symmetries 
of the particles. An example is the $n$-component version of the one-component $p-ip$ state discussed 
at the end of Sec.\ \ref{ssec:centralcharge}. The CFT consists of $n$ chiral Majorana fields, and has 
SO($n$) symmetry and currents, which give rise to an SO($n$) level-1 Chern-Simons term in the induced 
action, while the particle system may have U($n$) symmetry. This makes sense if we understand 
the trial state as a paired superfluid of $n$ particle types (each type pairing with itself), 
and the pairing as breaking the symmetry down from U($n$) to an SO($n$) subgroup. The system possesses 
non-Abelian Hall-conductivity responses in the unbroken SO($n$) subalgebra.

It might be interesting to pursue further extensions to conformal-block trial states in which the CFT has 
higher-spin current algebras, such as $W$- and superconformal algebras\cite{byb}, for which there are 
sometimes further coefficients that we expect can be obtained from Berry curvature. These algebras should 
correspond to additional structures on the spacetime, and to additional terms in the induced action.

\section{Conclusion}\label{sec:conclusion}

To summarize, we have shown that the topological central charge can be obtained from the Berry curvature
arising from the ground state wavefunction when the spatial metric is varied adiabatically. This method
can be applied in a numerical calculation. We showed analytically that it does produce the
expected result, equal to the central charge of the underlying CFT, when applied to topological phases
that possess conformal-block trial wavefunctions in which the generalized screening
hypothesis holds. We emphasized that this is a
{\em bulk} approach, which does not involve an edge, and that the result is a topological invariant that
characterizes a topological phase. More precisely, in some cases the central charge is obtained only in
combination with another topological property $\nu\overline{s^2}$, as emphasized in Sec.\ \ref{sec:disc}. We
explained how in general, if the value of the topological central charge is assumed known from the
edge theory, then the Berry curvature yields instead the value of $\nu\overline{s^2}$.

\acknowledgments

We would like to thank J. Dubail and G. Moore for helpful discussions, and we are grateful for the
hospitality of the Simons Center for Geometry and Physics in Stony Brook, NY, and of the Erwin 
Schr\"{o}dinger Institute in Vienna, Austria, at each of which parts of this work were conceived. We are 
also grateful to an anonymous referee, who suggested that we give more details for multicomponent states. 
The work of B.B. and N.R. was supported by NSF grants, Nos.\ DMR-1005895 and DMR-1408916.

\bibliography{berry}

\end{document}